\documentclass[10pt,lettersize,journal]{IEEEtran}
\usepackage[top=1.0in, bottom=1.04in, left=1.01in, right=1.01in]{geometry}
\usepackage{amsmath,graphicx,mathrsfs}
\usepackage{etex}
\usepackage{makecell}
\usepackage{float,subfigure}
\usepackage{cite}

\usepackage{bm}
\usepackage{times,amsmath,bm,epsfig,amssymb,algorithm}
\usepackage{booktabs}
\usepackage{algorithm} 
\usepackage{algorithmic}
\usepackage[algo2e]{algorithm2e} 
\usepackage[table ]{ xcolor}

\usepackage{xcolor}
\begin{document}
%
\title{VARFVV: View-Adaptive Real-Time Interactive Free-View Video Streaming with Edge Computing}

\author{
Qiang~Hu, \textit{Member, IEEE},~Qihan~He,~Houqiang~Zhong, \textit{Student Member, IEEE},~Guo~Lu, \textit{Member, IEEE},~Xiaoyun~Zhang, \textit{Member, IEEE},~Guangtao~Zhai, \textit{Fellow, IEEE}, ~Yanfeng~Wang
\thanks{
Manuscript received 15 May 2024,  revised 12 December 2024,  accepted 15 January 2025.
This work was supported in part by the National Key R\&D Program of China under Grant 2021YFE0206700,   STCSM (24511106902,  24ZR1432000,  24511106900),  National Natural Science Foundation of China (62271308,  62471290,  62331014),  111 plan (BP0719010),  Open Project of National Key Laboratory of China (23Z670104657) and State Key Laboratory of UHD Video and Audio Production and Presentation. \textit{(Corresponding authors: Guo Lu and Xiaoyun Zhang)}

Qiang Hu and Xiaoyun Zhang are with the Cooperative Medianet Innovation Center,  Shanghai Jiao Tong University,  Shanghai,  200240,  China.(e-mail: qiang.hu@sjtu.edu.cn;xiaoyun.zhang@sjtu.edu.cn)

Qihan He is with the School of Information Science and Technology,  ShanghaiTech University,  Shanghai,  201210,  China. (e-mail: heqihan1@alumni.shanghaitech.edu.cn)

Houqiang Zhong,  Guo Lu, and Guangtao Zhai are with the School of Electronics Information and Electrical Engineering,  Shanghai Jiao Tong University,  Shanghai,  200240,  China.
(e-mail: zhonghouqiang@sjtu.edu.cn; luguo2014@sjtu.edu.cn; zhaiguangtao@sjtu.edu.cn)

Yanfeng Wang is with the School of Artificial Intelligence, Shanghai Jiao Tong University,  Shanghai,  200240,  China. (e-mail: wangyanfeng@sjtu.edu.cn)

}
}

\maketitle
\pagestyle{empty}
\thispagestyle{empty}

\begin{abstract}

Free-view video (FVV) allows users to explore immersive video content from multiple views. However, delivering FVV poses significant challenges due to the uncertainty in view switching, combined with the substantial bandwidth and computational resources required to transmit and decode multiple video streams, which may result in frequent playback interruptions. Existing approaches, either client-based or cloud-based, struggle to meet high Quality of Experience (QoE) requirements under limited bandwidth and computational resources. To address these issues, we propose VARFVV, a bandwidth- and computationally-efficient system that enables real-time interactive FVV streaming with high QoE and low switching delay.
Specifically, VARFVV introduces a low-complexity FVV generation scheme that reassembles multiview video frames at the edge server based on user-selected view tracks, eliminating the need for transcoding and significantly reducing computational overhead. This design makes it well-suited for large-scale, mobile-based UHD FVV experiences. Furthermore, we present a popularity-adaptive bit allocation method, leveraging a graph neural network, that predicts view popularity and dynamically adjusts bit allocation to maximize QoE within bandwidth constraints. We also construct an FVV dataset comprising 330 videos from 10 scenes, including basketball, opera, etc. Extensive experiments show that VARFVV surpasses existing methods in video quality, switching latency, computational efficiency, and bandwidth usage, supporting over 500 users on a single edge server with a switching delay of 71.5ms. Our code and dataset are available at  https://github.com/qianghu-huber/VARFVV.

\end{abstract}

\IEEEpeerreviewmaketitle

\begin{figure*}
  \includegraphics[width=0.98\linewidth]{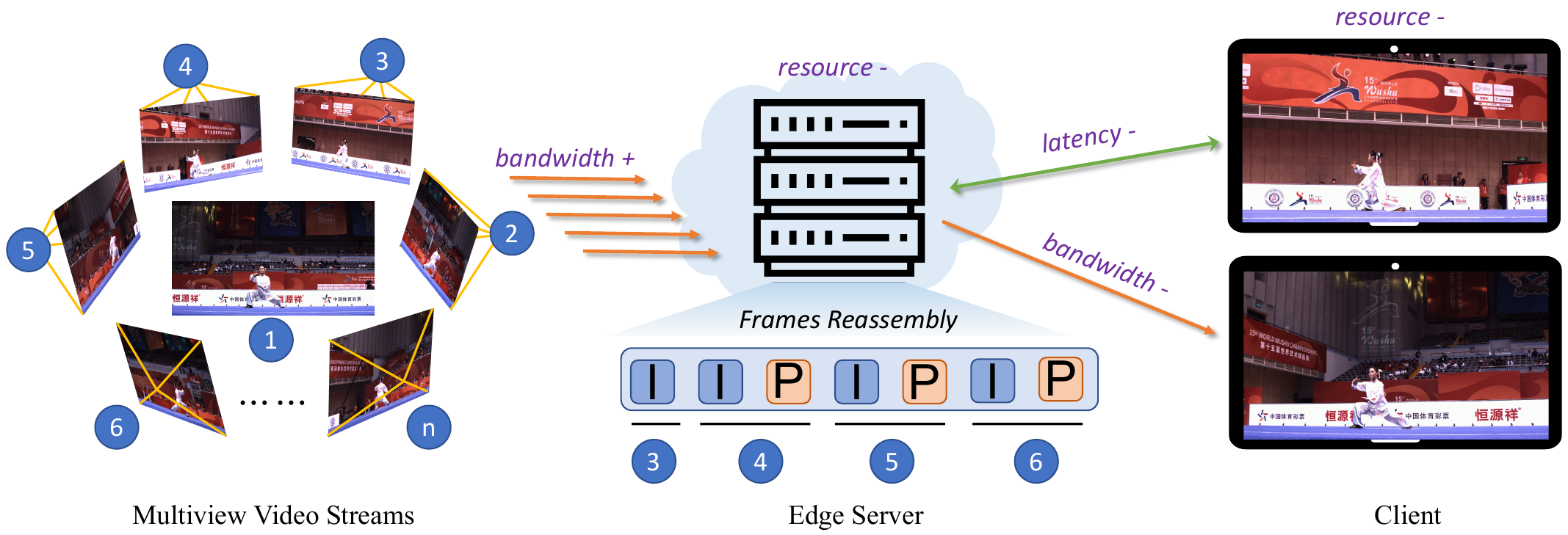}
  \caption{Demonstration of our VARFVV. VARFVV is a bandwidth-efficient and low-complexity interactive FVV streaming system and employs an edge server to reassemble multiview frames continuously in time and inter-view based on the user's view track to obtain the FVV stream. VARFVV further delivers the generated FVV stream to the user with low latency through WebRTC. This approach significantly reduces transmission bandwidth and computational resources needed at the edge server and client.
  }
  \label{fig:teaser}
\end{figure*}

\section{Introduction}

\IEEEPARstart{I}{nteractive} free-view video (FVV) is an emerging technology that allows users to freely choose their viewing angle as if they were present in the scene. FVV captures dynamic 3D scenes using multiple closely spaced, time-synchronized cameras, encoding the data into multiple video streams to provide panoramic scenes.  This technology enables immersive live streaming with smooth inter-view switching and dynamic bullet-time effects, making it highly promising for large-scale broadcasts such as concerts, sports, and interactive teaching.

Delivering FVV over mobile networks faces several challenges regarding video quality, view switching delays, transmission bandwidth, and computational resources, which are often in conflict with one another. Achieving high-quality video requires increased bandwidth and processing power to manage tasks such as loading, decoding, and rendering video streams. However, the inherent limitations of mobile networks and devices restrict their capacity to provide sufficient bandwidth and processing power, particularly in bandwidth-limited or resource-constrained environments. As video quality improves (e.g., higher bitrate or resolution), the bandwidth needed for transmission increases proportionally, which can result in network congestion and playback interruptions. Additionally, frequent view switching demands real-time data loading, decoding, and rendering, imposing substantial computational strain on both servers and mobile devices and often leading to latency and degraded user experience. Efforts to optimize one aspect, such as reducing bandwidth usage or minimizing switching delays, may inadvertently compromise video quality or increase computational demands, thereby intensifying performance issues. These intricate trade-offs underscore the necessity of developing effective strategies to balance conflicting factors and to ensure seamless and efficient delivery of FVV over mobile networks.

Early methods \cite{fvtv1,fvtv2} for FVV systems typically involve transferring all views to the client, either encoded together or separately. Although these approaches ensure that all views are available, they are highly inefficient, leading to excessive bandwidth consumption and high decoding burdens on the client side. Such inefficiencies often degrade the Quality of Experience (QoE), particularly in bandwidth-constrained or resource-limited scenarios. In response, more recent solutions, such as HTTP Adaptive Streaming (HAS) or DASH-based techniques \cite{7406698, 8325530, 8480858, 8907860}, dynamically fetch some videos only adjacent to the current view, instead of all views, to reduce bandwidth usage. The video client, however, needs to clear the current buffer and re-buffer a fixed number of new frames before playback resumes when a user continuously and rapidly switches views. The delay in resuming video playback often has a negative impact on the viewing experience.

 Cloud-based delivery systems \cite{9431586, 9780408} have been developed to optimize transmission bandwidth and reduce client-side computational complexity in FVV streaming. These systems decode multiple video streams at the edge server and employ dedicated encoders to transcode the user-selected views, effectively minimizing bandwidth consumption and ensuring smooth view switching. However, as the number of users grows, the reliance on individual encoders for each user substantially increases the computational load on the server, leading to scalability and resource challenges. For example, an NVIDIA RTX4000 GPU with a hardware-based encoder (NVENC) can encode approximately 20 videos at 1080P@25FPS or 5 videos at 4K@25FPS in real-time simultaneously\footnote{https://developer.nvidia.com/nvidia-video-codec-sdk}. Supporting 500 users at 1080P@25FPS would require approximately 25 RTX4000 GPUs. Therefore, it is essential to explore more efficient strategies for FVV live systems.

\IEEEpubidadjcol

In this paper, we propose a novel view-adaptive real-time interactive FVV streaming system with edge computing or VARFVV, which achieves low switching delay and high QoE while maintaining low-cost computation and transmission, as illustrated in Fig. \ref{fig:teaser}. We realize this through two main innovations. First, we design a low-complexity FVV generation scheme that reassembles multiview video frames at the edge server based on user-selected view tracks, avoiding the need for transcoding. This reassembly approach greatly accelerates the generation process compared to traditional transcoding-based cloud methods. The generated FVV stream is then delivered to users in real-time through WebRTC \cite{WebRTC}, ensuring minimal latency. By reassembling and transmitting a single video stream to the client, VARFVV significantly reduces both transmission bandwidth consumption and the computational load on the server and client, enabling the system to efficiently support a large number of users simultaneously experiencing high-quality UHD FVV live services on mobile devices.

 \begin{table*}[ht]
\centering
\caption{Comparison of FVV streaming methods across key dimensions. Our VARFVV achieves low view switching latency and bandwidth usage, high video quality, and minimal resource usage at both the client and edge server.}
\label{comparison FVV}
\resizebox{\linewidth}{!}{%
\begin{tabular}{lcccccc}
\toprule
\textbf{Method} & \textbf{View Switching Latency}$\downarrow$  & \textbf{Video Quality}$\uparrow$ & \textbf{Bandwidth Usage}$\downarrow$ & \makecell{\textbf{Client} \\ \textbf{Computing Resource}}$\downarrow$ & 
\makecell{\textbf{Edge Server} \\ \textbf{Computing Resource}}$\downarrow$ \\ 
\midrule

HASFVV \cite{8325530} & High & Low & High & High & - \\

Yao \textit{et al.} \cite{8480858} & High & Low & High & High & - \\

DASHFVV [23] & High & Low & High & High & - \\


EdgeEncodingFVV \cite{9431586} & Low & High & Low & Low & High \\

HybridFVV \cite{9780408} & Low & High & Low & Low & High \\

Hu \textit{et al.} [31] & Low & High & Low & Low & High \\
\rowcolor{gray!20} \bf VARFVV (Ours) & \bf Low &  \bf High & \bf Low & \bf Low & \bf Low \\ 
\bottomrule
\end{tabular}%
}
\end{table*}

Second, we introduce a popularity-adaptive bit allocation scheme to further enhance the overall QoE within a constrained bit budget. Our approach begins by formulating a QoE-aware optimization problem aimed at maximizing the user's QoE. To solve this problem, we introduce a graph neural network (GNN) based view popularity prediction method, where each view is modeled as a vertex within VARFVV. Transitions between vertices are used to characterize users' view trajectories, allowing us to estimate the popularity of each view. Based on this prediction, we design a novel bit allocation scheme that dynamically allocates more bits to perceptually popular views by borrowing bits from less popular ones. This adaptive approach optimizes the video quality of the most viewed perspectives, ensuring that the system maximizes the perceived visual quality for users and effectively balances bit resources to enhance the overall QoE.

In summary, our contributions are as follows:

\begin{enumerate}
\item[$\bullet$] We propose VARFVV, a bandwidth- and computationally-efficient system that enables real-time interactive FVV streaming with high QoE and low switching delay, providing a scalable solution for delivering immersive FVV services in real-world scenarios.

\item[$\bullet$] We introduce a low-complexity FVV generation scheme that reassembles multiview video frames instead of transcoding, significantly reducing the computational load on the edge server. Experimental results show that a single edge server with an AMD Ryzen 7 3700 CPU@3.6 GHz can support over 500 users simultaneously experiencing FVV.

\item[$\bullet$] We develop a spatial-temporal graph network to predict view popularity and implement a popularity-based bit allocation algorithm, maximizing overall QoE by prioritizing frequently viewed perspectives.

\end{enumerate}

\section{Related Work}
\subsection{Interactive Free-View Video Streaming} \label{r1}

Early methods \cite{fvtv1,fvtv2} achieve free-view video by transferring all video streams to a client. Even though the multiview videos can be compressed by exploiting the dependencies\cite{4358661, 4358662}, transmitting all videos to a client is not optimal because the client only uses a small portion of them. Receiving all video streams also results in high bandwidth and consumption for the client. Some methods \cite{5557810,4665121} employ redundant coding structures, where each original frame is encoded into multiple versions, appropriately trading off expected transmission rate with storage, to facilitate view switching. However, in scenarios where clients frequently switch views, both methods require re-buffering a fixed number of new frames to resume playback.

To enhance bandwidth efficiency, adaptive streaming techniques \cite{7406698, 7944625, 7875099, 8907860, 10.1016/j.comcom.2016.04.001, 7862898, 10.1145/2733373.2807971, 10.1145/2910017.2910610, 8480858, 9148582, 10.1145/3204949.3204968, 30, 8325530,8806082} have been proposed for multiview and free-view video systems. These techniques enable the transmission of videos with varying numbers and bitrates. For instance, Zhang \textit{et al.} \cite{8325530} address the limitation of equal bitrate by jointly optimizing reference views and bitrates to select the optimal subset of different views and bitrates. A DASH-based FVV streaming client is proposed to reduce view switching latency by prefetching a set of adjacent images \cite{8806082}. However, even with the simultaneous use of 9 decoders, the view switching time remains at least 800 ms. With the advancement of deep learning technologies, several methods \cite{9605672,10368051,9996351,10158528,10033423,9693960,9953110} have leveraged deep learning-guided resource multiplexing and view semantic information to further enhance bandwidth efficiency. Although the adaptive streaming approaches partially reduce the bandwidth and client-side computation, they still need to receive and decode redundant views.

Cloud-based delivery systems \cite{9431586, 9780408,9859922} offer a promising solution by transferring bandwidth and computational pressure to the server. In these systems, the edge server uses dedicated encoders to transcode the user-selected view into an FVV stream, transmitting only the requested target stream to the client. For example, Dong \textit{et al.} \cite{9431586} propose an elastic system architecture with low-latency features to support generic interactive video applications at near-user edges by extending the directed acyclic graph model with switch vertices. However, when the number of users is large, the edge server may face significant computing resource demands for transcoding. In contrast, our method focuses on reassembling video frames instead of transcoding them at the edge, thus maintaining the original video quality without imposing heavy computational demands.
Tab. \ref{comparison FVV} presents a comparative analysis of FVV streaming methods across key performance metrics: view switching latency, video quality, bandwidth usage,  client computing resources, and edge server computing resources. The proposed VARFVV achieves low view switching latency and bandwidth usage while maintaining high video quality with minimal resource requirements for both the client and the edge server.

\begin{figure*}[ht]
\includegraphics[width=0.97\linewidth]{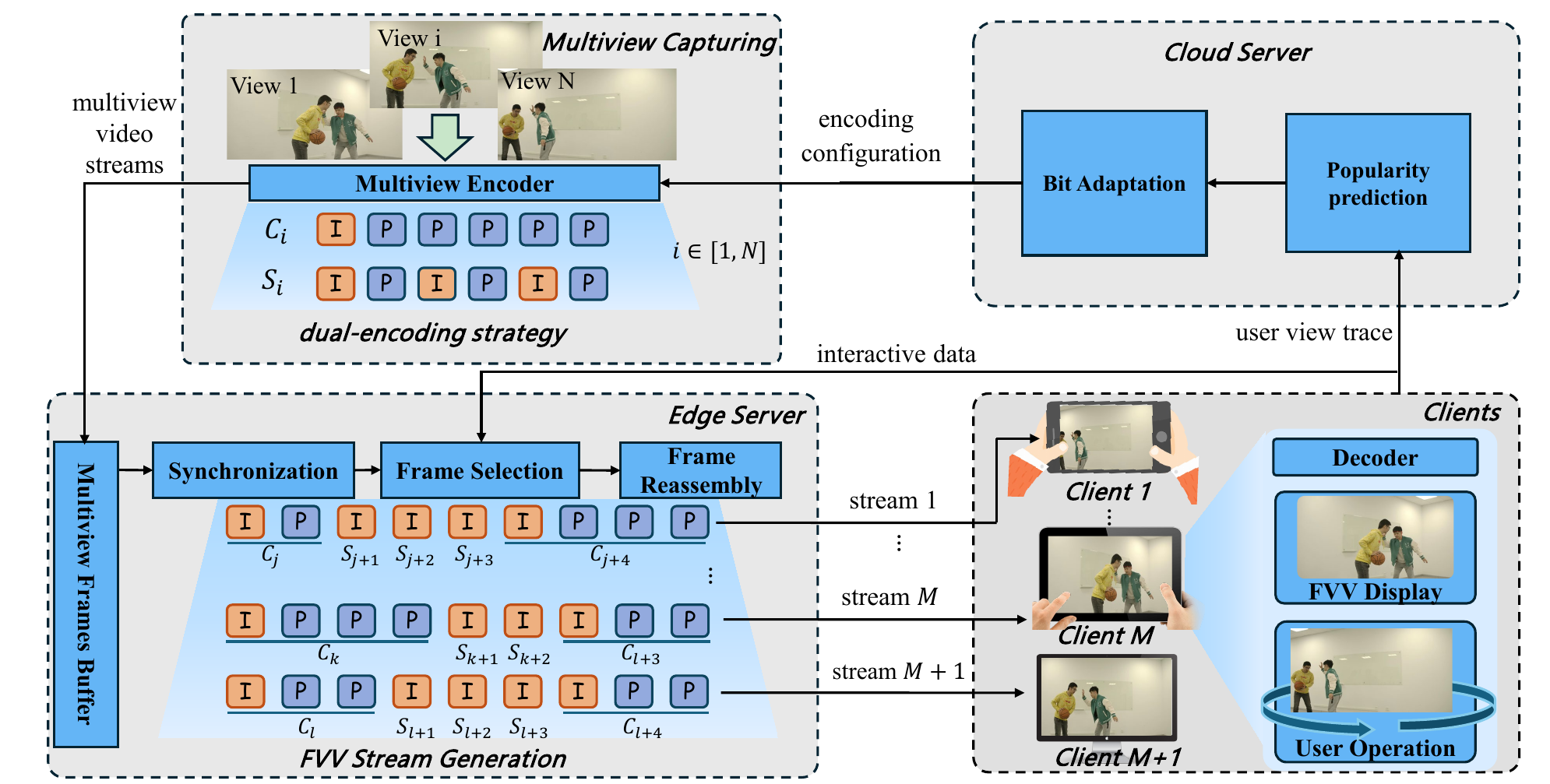}
\caption{Overview of the VARFVV System Architecture. Our VARFVV integrates multiview encoding, popularity-adaptive bit allocation, edge-side FVV stream generation, and client-side decoding to deliver high-quality, real-time FVV services with low computational and transmission costs.}
\label{overview}
\end{figure*}

\subsection{Bit Allocation}
Optimal bit allocation schemes for video coding with a given visual attention have been investigated in \cite{ bottom2_coding, roi1_coding, C_W1, C_W2, Zhou_Wang, visual_saliency, ICIP2016, Guo_tip, hu2020loop, Jeong2023Delta}. By allocating more bits to attended or popular regions, it is possible to design video coding algorithms that can improve visual quality while meeting bandwidth constraints. Gitman \textit{et al.} \cite{roi1_coding} develop a gaze-based video coding framework to achieve high-quality salient regions. For 360$^\circ$ video delivery, some work \cite{streaming1,streaming2,streaming3, Chen2023Live360,Huo2022TS360AT,Chen2024MacrotileTQ} proposes viewport-adaptive 360$^\circ$ video streaming systems to reduce the bandwidth waste.  Chen \textit{et al.} \cite{Chen2023Live360} introduce a viewport-aware upstream and downstream bit allocation algorithm to maximize users’ QoE.

Accurately predicting view popularity is important for determining the optimal bit allocation strategy. Chakareski \textit{et al.} \cite{vpd} present a view popularity-aware bit allocation algorithm to maximize the expected video quality, but their popularity prediction method only relies on statistical data and ignores temporal characteristics. To address this, some studies use the Long Short-Term Memory (LSTM) network to predict future video popularity based on historical data \cite{make10.1145/3310165.3310174, Zink2019, 9343267, Cheng2022,9567690}. For example, Maniotis \textit{et al.} \cite{9343267} employ an LSTM network to forecast the popularity evolution of the content requests and prefetch content for caching. However, LSTM-based methods focus primarily on temporal patterns, often neglecting spatial relationships among tiles that are critical for accurate predictions in 360$^\circ$ video.

Recent advancements in graph-based modeling for 360$^\circ$ video have demonstrated superior performance in predicting user viewport changes by integrating spatial and temporal dependencies. Hu \textit{et al.} \cite{hu2020tvg} model changes in user viewport trajectories as a directed weighted graph, leveraging the tile transition probability matrix to predict tile view popularity. Similarly, Xu \textit{et al.} \cite{gnn360streaming} model user viewport transitions using a graph structure and further incorporates GNNs to fuse multiple features for predicting tile popularity. These graph-based approaches capture spatial and temporal dependencies more effectively, significantly improving the accuracy of popularity predictions and enabling efficient resource allocation in 360$^\circ$ video streaming systems. In this paper, we extend the concept of graph-structured modeling from 360$^\circ$ video to FVV by representing view transitions as a weighted directed graph. Leveraging a GNN-based popularity prediction algorithm and a popularity-adaptive bit allocation strategy, our approach integrates spatial and temporal dynamics to maximize QoE under bandwidth constraints.

\section{System Design and Problem Formulation}

\subsection{System Overview}

The VARFVV system is designed to deliver high-quality, real-time FVV services in a cost-effective manner by integrating multiview encoding, popularity-adaptive bit allocation, edge-side FVV stream generation, and client-side decoding within a unified architecture, as shown in Fig. \ref{overview}. The system captures dynamic 3D scenes from multiple views using a synchronized array of ZCAM\footnote{http://www.z-cam.com/e2/} cameras, and the captured views are then encoded via the ``\textbf{Multiview Encoder}" using the H.264 codec through the FFmpeg\footnote{https://ffmpeg.org/} library. Each view ($v_i$) is encoded into two formats: \emph{view-switching representation} ($S_i$) for rapid view changes and \emph{view-constant representation} ($C_i$) for stable perspectives. This dual-encoding approach forms the foundation for continuous and seamless view switching.

To optimize the QoE for users, the system employs a popularity-adaptive bit allocation algorithm on the cloud server. The ``\textbf{Popularity Prediction}" module analyzes limited historical user interaction data to forecast the future popularity of various views. According to the predicted popularity of each view, the ``\textbf{Bit Adaptation}" component periodically updates the bit allocation of the multiview encoder. By borrowing bits more aggressively from perceptually less important views and allocating them to perceptually popular views, we improve the overall quality of FVV transmitted to users while meeting bandwidth constraints.

The edge-side FVV generation scheme is crucial for minimizing view switching delays and reducing transmission and computational costs.
The ``\textbf{Synchronization}" module aligns video frames from multiple views using their Presentation Time Stamps (PTS) to ensure accurate synchronization. When users interact with the system, such as swiping or switching views, the ``\textbf{Frame Selection}" and ``\textbf{Frame Reassembly}" components dynamically select and reassemble the appropriate frames into an FVV stream based on the user’s chosen trajectory. Since our system only reassembles the video frames rather than performing transcoding, it maintains low computational overhead.

The generated FVV stream is transmitted to the client using WebRTC technology, which guarantees low-latency delivery and immediate response to user interactions. The client-side decoding module processes the incoming stream locally, sending interaction signals to the server as required. With the VARFVV design, our system efficiently enables users to enjoy high-quality UHD FVV live services on mobile devices over mobile networks. Detailed implementation specifics are available in Section \ref{sec:imple}. Tab.\ref{notation} summarizes the main notations used in the following sections.

\begin{table}[thbp]
\centering
\caption{Notation}
\small
  \begin{tabular}{ccl}
    \toprule
    Symbol & Physical Meaning\\ 
    \midrule
     \texttt{$v_{i}$}& The \emph{i}-th view\\
    \texttt{$S_{i}$}&  The \emph{view-switching representation} of $v_i$\\
    \texttt{$C_{i}$}& The \emph{view-constant representation} of $v_i$\\
    \texttt{$S_{i,j}$}& The \emph{j}-th video chunk in $S_{i}$ \\
    \texttt{$C_{i,j}$}& The \emph{j}-th video chunk in $C_{i}$ \\
    \texttt{$x_{i,j}$}& Actual popularity of $C_{i,j}$\\ 
    \texttt{$\hat{x}_{i,j}$}& Actual popularity of $S_{i,j}$\\ 
    \texttt{$p_{i,j}$}& Predicted popularity of $C_{i,j}$\\ 
    \texttt{$\hat{p}_{i,j}$}& Predicted popularity of $S_{i,j}$\\ 
    \texttt{$R_{i,j}$}& Allocated bits for $C_{i,j}$\\ 
    \texttt{$\hat{R}_{i,j}$}& Allocated bits for $S_{i,j}$ \\
    \bottomrule 

  \end{tabular}
  \label{notation}
 \end{table}

\subsection{FVV Stream Generation} \label{gop}

In traditional cloud-based delivery systems \cite{9431586, 9780408,9859922}, each user request triggers re-encoding of the video stream on the server, leading to significant computational overhead. To alleviate this, we propose a low-complexity FVV stream generation scheme that reassembles the I-frames and P-frames from the multiview video stream at the edge server based on user-selected view paths. In video coding, I-frames are independently decodable, while P-frames rely on previous I/P-frames for decoding \cite{h264, hevc}. Consequently, view switching requires waiting for the next view’s I-frame, with longer I-frame intervals causing delays that degrade the user experience. To enable random access for low-latency view switching, multiview videos can be encoded with high-density I-frames at the acquisition stage. However, as I-frames require more bits than P-frames to achieve the same quality, transmitting high-density I-frames results in wasted bandwidth when the user's view is constant.

  \begin{figure}[t]
\centerline{\includegraphics[width=\linewidth]{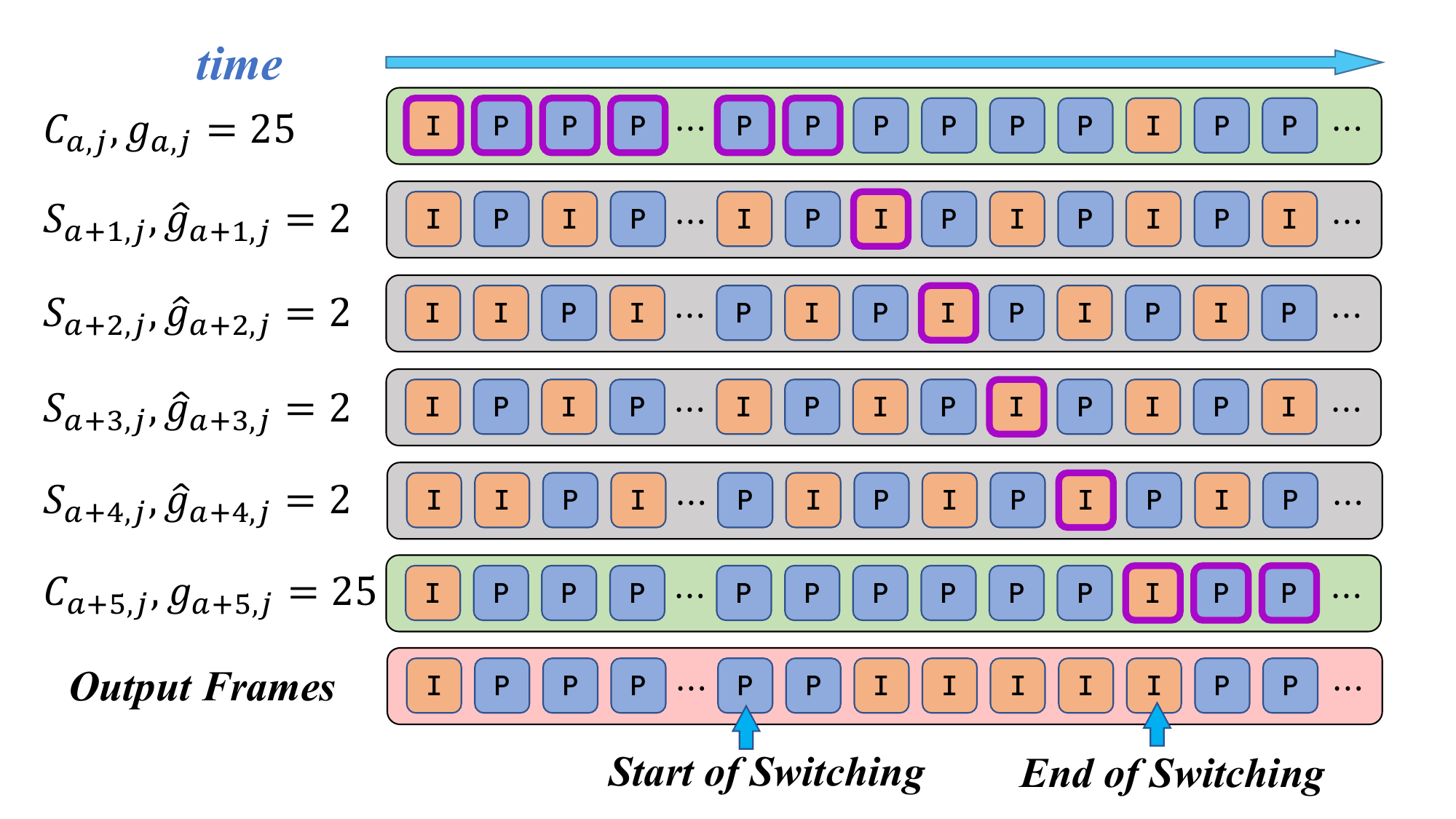}}
\caption{GoP structure design for each stream and view switching strategy. }
\label{view-switching}
\end{figure}

 To address this issue, we introduce a dual-encoding strategy at the acquisition stage, where each view $v_i$ ($1\leq i\leq N$) is encoded into two representations: a \emph{view-switching representation} labeled as $S_i$ and a \emph{view-constant representation} labeled as $C_i$, as shown in Fig. \ref{view-switching}. The former uses a static group of pictures (GoP) size 2 ($\hat{g}_{i,j}=2$) and is used for view-switching scenarios, while the latter uses a static GoP size 25 ($g_{i,j}=25$) and is used for view-constant scenarios. The encoded multiview streams are transmitted to the edge server, where they are dynamically reassembled based on the user’s view paths. When a user's current view is fixed at $v_i$ for the $j$-th video chunk, the edge server extracts the compressed video frames directly from $C_{i,j}$ as the output frames and transmits them to the user. If the user switches the view from $v_i$ to $v_m$, the edge server immediately selects a time-continuous I-frame from $S_{i+1,j}$ to $S_{m,j}$ and reassembles these frames consecutively in time to generate the FVV stream transmitted to the user. After view switching is completed, the edge server continues to multiplex the video frames in $S_{m,j}$ as the output frame until it encounters a time-synchronized I-frame in $C_{m,j}$, at which point it starts to reuse the frames in $C_{m,j}$. This process enables continuous, seamless, and fast switching of user views with an almost negligible increase in computational complexity. It is worth noting that when $i$ is an even number, we insert an I-frame at the starting position of $S_{i,j}$ so that the I-frames between two adjacent \emph{view-switching representations} are staggered, which effectively reduces the switching delay.

\subsection{QoE Metrics}

We aim to maximize the QoE with limited bandwidth resources by determining the bit allocation for each view at a specific video chunk $j$. $R_{i,j}$ and $\hat{R}_{i,j}$ denote the allocated bits for the $j$-th video chunk in $C_i$ and $S_i$, respectively, where $i\in\{1,2,\cdots,N\}$ is the view number. We model the total QoE of FVV for a video chunk $j$ as follows:
\begin{align}
    & \mathbf{QoE_j}(R_{1,j},\dots,R_{N,j},\hat{R}_{1,j},\dots,\hat{R}_{N,j}) = \nonumber \\
    & \mathbf{QoE_{1,j}}  -  \mu_1 \mathbf{QoE_{2,j}}  - \mu_2 \mathbf{QoE_{3,j}}
\end{align}
where $\mathbf{QoE_{1,j}}$ denotes the QoE on video quality, $\mathbf{QoE_{2,j}}$ and $\mathbf{QoE_{3,j}}$ represent inter-view and temporal quality switching, respectively. These three QoE features are essential to characterizing the QoE of FVV. $\mu_1$ and $\mu_2$ are weighted factors.

\textbf{Video Quality.}
 Following the previous work \cite{ 7564469, hu2020tvg},  we adopt a logarithmic model to depict the relationship between QoE and allocated bits.
\begin{equation}
 \mathbf{QoE_{1,i,j}} = log(1+R_{i,j}/\eta)
\end{equation}
where $\eta$ is a model parameter. Considering all  \emph{view-switching representations} and  \emph{view-constant representations}, we obtain the QoE on video quality for all views as follows:
\begin{equation}
 \mathbf{QoE_{1,j}} = \sum_{i=1}^{N} x_{i,j}log(1+R_{i,j}/\eta )+\sum_{i=1}^{N} \hat{x}_{i, j}log(1+\hat{R}_{i,j}/\hat{\eta} )
 \label{eq:qoe1}
\end{equation}
where $x_{i, j}$ and $\hat{x}_{i, j}$ are the view popularity of $C_i$ and $S_i$ in the $j$-th video chunk, respectively. $\hat{\eta}$ is a model parameter. As $x_{i, j}$ and $\hat{x}_{i, j}$ are unknown a prior, we propose a novel approach in Section \ref{GNN} that leverages historical view requests to obtain the predited popularity $p_{i,j}$ and $\hat{p}_{i,j}$. Then Eq. \ref{eq:qoe1} can be written as:
\begin{equation}
 \mathbf{QoE_{1,j}} = \sum_{i=1}^{N} p_{i,j}log(1+R_{i,j}/\eta )+\sum_{i=1}^{N} \hat{p}_{i, j}log(1+\hat{R}_{i,j}/\hat{\eta} ).
 \label{eq:qoe2}
\end{equation}

\textbf{Inter-view Quality Switching.}
When users switch views in an FVV system, they encounter variations in video quality due to the transitions between different view representations, each potentially encoded with varying bitrates. These transitions can occur between two consecutive \emph{view-switching representations} or between a \emph{view-constant representation} and a \emph{view-switching representation}.  Excessive bitrate changes between two consecutive representations can potentially lead to degraded visual quality when switching views. To measure and control these variations, we define the inter-view quality switching metric as follows:
\begin{equation}
\begin{aligned}
 \mathbf{QoE_{2,j}} = \mu_3 \sum_{i=2}^{N}\hat{p}_{i,j}(\hat{R}_{i,j}-\hat{R}_{i-1,j})^2+ \\ \sum_{i=2}^{N}\hat{p}_{i,j}(\hat{R}_{i,j}/\hat{\eta}-R_{i-1,j}/\eta)^2    
\end{aligned}
\end{equation}
where $\mu_3$ is a weighted factor. The first term in the equation quantifies the quality variation between consecutive \emph{view-switching representations}, calculating the squared difference in bitrate between two successive view-switching events, weighted by the predicted popularity $\hat{p}_{i,j}$. The second term evaluates the transition between a \emph{view-constant representation} and a \emph{view-switching representation}. The metric $ \mathbf{QoE_{2,j}}$ is designed to capture the magnitude of bitrate changes between consecutive representations, enabling the system to monitor and regulate these fluctuations, thus ensuring visual consistency and improving the overall QoE for users.

\textbf{Temporal Quality Switching.}  Temporal quality switching refers to the variation in video quality between consecutive video chunks when the user’s viewing angle remains fixed, as represented in the \emph{view-constant representation} $C_i$. Significant changes in the bitrate between these consecutive chunks can cause discomfort or adverse physiological effects, such as dizziness or motion sickness, particularly in immersive environments like FVV. To accurately quantify these variations, we define the temporal quality switching metric as follows:
\begin{equation}
 \mathbf{QoE_{3,j}} = \sum_{i=1}^{N}p_{i,j}(R_{i,j}-R_{i,j-1})^2.
\end{equation}
where $R_{i,j}$ and $R_{i,j-1}$ denote the bitrate of the current and previous video chunks, respectively. The squared difference between the bitrates of consecutive video chunks reflects the magnitude of the variation,  with larger fluctuations penalized more heavily. This design prioritizes video quality stability over time, reducing the risk of visual inconsistency.

\subsection{Optimization Problem Formulation}

We can now formulate a QoE-aware optimization problem for FVV adaptive bit allocation.
\begin{equation}
\begin{aligned}
& \mathbf{arg  \ \   max} \ \   \mathbf{QoE_j}(R_{1,j},\dots,R_{N,j},\hat{R}_{1,j},\dots,\hat{R}_{N,j}) \\
& \mathbf{s.t.} \ \ \sum_{i=1}^{N} R_{i,j} + \hat{R}_{i,j} \le R_j \\ & \quad \quad R_{\mathbf{min}} \le R_{i,j} \le R_{\mathbf{max}}, \hat{R}_{\mathbf{min}} \le \hat{R}_{i,j} \le \hat{R}_{\mathbf{max}} 
\end{aligned}
\label{problem}
\end{equation}

The objective of our adaption logic is to optimize the bit allocation $R_{i,j}$ and $\hat{R}_{i,j}$ ($i\in\{1,2,\cdots,N\}$) for a video chunk $j$, so that the total QoE measured on the $j$-th video chunk is maximized. The constraint restricts an upper bound on the sum of allocated bits $R_j$, and requires that the allocated bits $R_{i,j}$ and $\hat{R}_{i,j}$ should not fall below $R_{\mathbf{min}}$ and $\hat{R}_{\mathbf{min}}$, respectively, which correspond to the lowest available video quality. To prevent excessive use of bits, $R_{i,j}$ and $\hat{R}_{i,j}$ should also not exceed $R_{\mathbf{max}}$ and $\hat{R}_{\mathbf{max}}$. The optimal bit allocation problem in Eq. \ref{problem} can be broken down into two parts. The first is to predict the view popularity $p_{i,j}$ and $\hat{p}_{i,j}$, while the second is to assign bit based on the predicted view popularity. In Section \ref{sec:4}, we propose a tractable approach to solve Eq. \ref{problem}.

Unlike \cite{hu2020tvg}, which focuses on tile-based bitrate allocation for $360^\circ$ video streaming, our approach is tailored for bitrate allocation across multiple camera views in FVV streaming. This shift from tile-level optimization to inter-view transitions requires additional constraints and metrics specifically designed for dynamic view switching.

\section{Optimal Bit Allocation}
\label{sec:4}

\subsection{Popularity Prediction}
\label{GNN}

Here, we introduce a novel method based on historical data to predict view popularity $p_{i,j}$ and $\hat{p}_{i,j}$. For \emph{view-constant representations}, where users typically maintain a fixed viewing angle after selecting a preferred view, a baseline method leverages the ground truth popularity of each view ($x_{i,j-1}$) obtained from the previous video chunk as the predicted popularity of the next video chunk ($p_{i,j}$). We refer to this method as the previous popularity carryover (PPC) model.

However, for \emph{view-switching representations}, the popularity of views can change rapidly as users switch their focus between different views, resulting in highly nonlinear and complex patterns. Traditional methods like LSTM networks \cite{make10.1145/3310165.3310174, Zink2019} capture temporal dynamics from historical data, but they fail to account for the spatial relationships between views, which are crucial for accurate view popularity prediction in FVV. Inspired by recent works \cite{hu2020tvg, gnn360streaming} that model user viewport changes as directed weighted graphs in $360^\circ$ video tiles, we extend Guo \textit{et al.}'s attention-based spatial-temporal GNN approach \cite{guo2019attention}, originally designed for traffic intersection prediction, to FVV, enabling the capture of both temporal and spatial dependencies. 

\begin{figure}
    \centering
    \includegraphics[width=\linewidth]{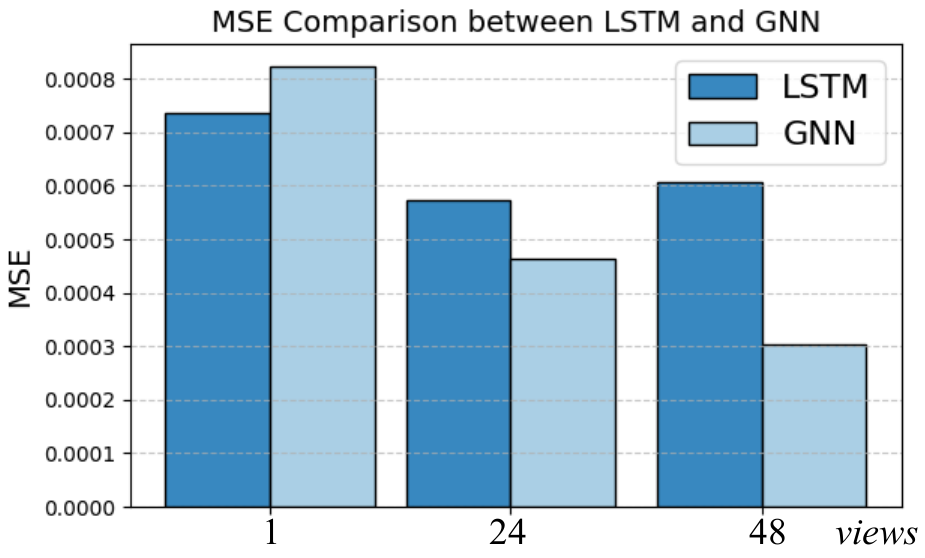}
    \caption{Comparison of prediction accuracy between LSTM and GNN methods for view popularity prediction across different numbers of views. }
    \label{fig:GNNLSTM}
\end{figure}

We preliminarily compare the performance of LSTM and GNN methods for predicting view popularity across varying numbers of views (single view, 24 views, and 48 views) in Fig. \ref{fig:GNNLSTM}. The results demonstrate that LSTM's accuracy does not improve with an increasing number of views, due to its limited ability to capture spatial relationships between views. In contrast, GNN’s graph-based modeling effectively addresses this issue, leading to more accurate predictions.
Our adapted GNN effectively captures both spatial correlations between views and temporal dynamics of user preferences, enabling accurate prediction of $\hat{p}_{i,j}$ for dynamic view-switching scenarios.

\begin{figure}[t]
    \centering
    \includegraphics[width=\linewidth]{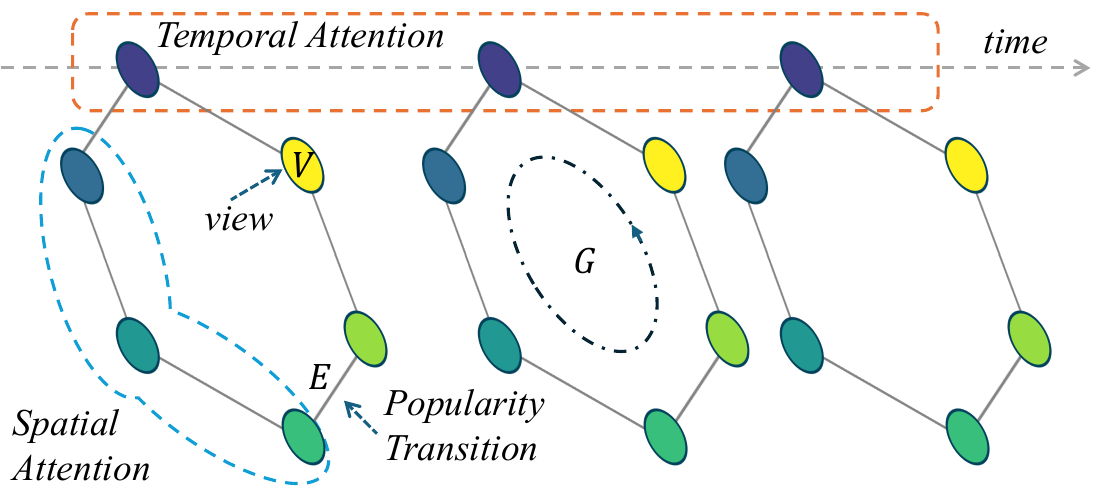}
    \caption{Graph-based model for view popularity prediction with spatial and temporal attention.}
    \label{fig:gnn-model}
\end{figure}

Our attention-based spatial-temporal GNN models each view as a vertex in the graph $G$, with the popularity of each view in a video chunk as an attribute associated with the vertex, as shown in Fig. \ref{fig:gnn-model}. An undirected edge is established between vertices representing views that are adjacent based on camera connectivity. The adjacency matrix $\mathbf{A}$ is constructed to reflect these camera connections, capturing the spatial relationships between views in the graph structure. This scheme can be expressed as:  
\begin{align}
    G &=(V,E,\mathbf{A}) \\
    |V| &= N, \mathbf{A} \in \mathbb{R}^{N \times N} \nonumber
\end{align}  
where $N$ is the total number of views, $V$ represents the set of vertices, $E$ denotes the edges based on the logical view switching  between cameras. For each frame, spatial attention is applied to aggregate information from neighboring nodes, while temporal attention aggregates data from the same node across different time steps. The attention mechanism enables the model to prioritize relevant spatial and temporal features, improving its ability to capture spatial correlations (e.g., adjacent views) and temporal dynamics (e.g., changes in user preferences). This design ensures that the model remains robust and effective, even in scenarios with sudden changes in user behavior, as verified in the experimental section.  

We represent the popularity of $S_{i}$ in the $j$-th chunk as $\hat{x}_{i,j} \in \mathcal{R}, i \in N$. 
$ \mathbf{X}_{j} = (\hat{x}_{1,j}, \hat{x}_{2,j} ,\cdots ,\hat{x}_{N,j}) $ denotes the popularity of all nodes for the $j$-th video chunk. 
$\mathcal{X}_{\tau}  = (\mathbf{X}_1, \mathbf{X}_2, \cdots, \mathbf{X}_{\tau}) $ represents the historical popularity of all nodes over $\tau$ chunks. 
As shown in Eq. \ref{eq:9}, graph neural network $F$ maps the historical popularity $\mathcal{X}_{\tau}$ to future view popularity sequences $\mathcal{P}^{d}$ over the next $d$ chunks.
$\mathcal{P}^{d}=(\mathbf{P}_{\tau+1}, \mathbf{P}_{\tau+2} ,\cdots ,\mathbf{P}_{\tau+d})$, where $\mathbf{P}_{\tau+1}=(\hat{p}_{1,\tau+1}, \hat{p}_{2,\tau+1} ,\cdots ,\hat{p}_{N,\tau+1})$.
\begin{align}
 \begin{aligned}
    \mathcal{P}^{d} = &F(\mathcal{X}_{\tau}) \label{eq:9} \\
\end{aligned}       
\end{align}

\begin{figure}[t]
    \centering
    \includegraphics[width=1\linewidth]{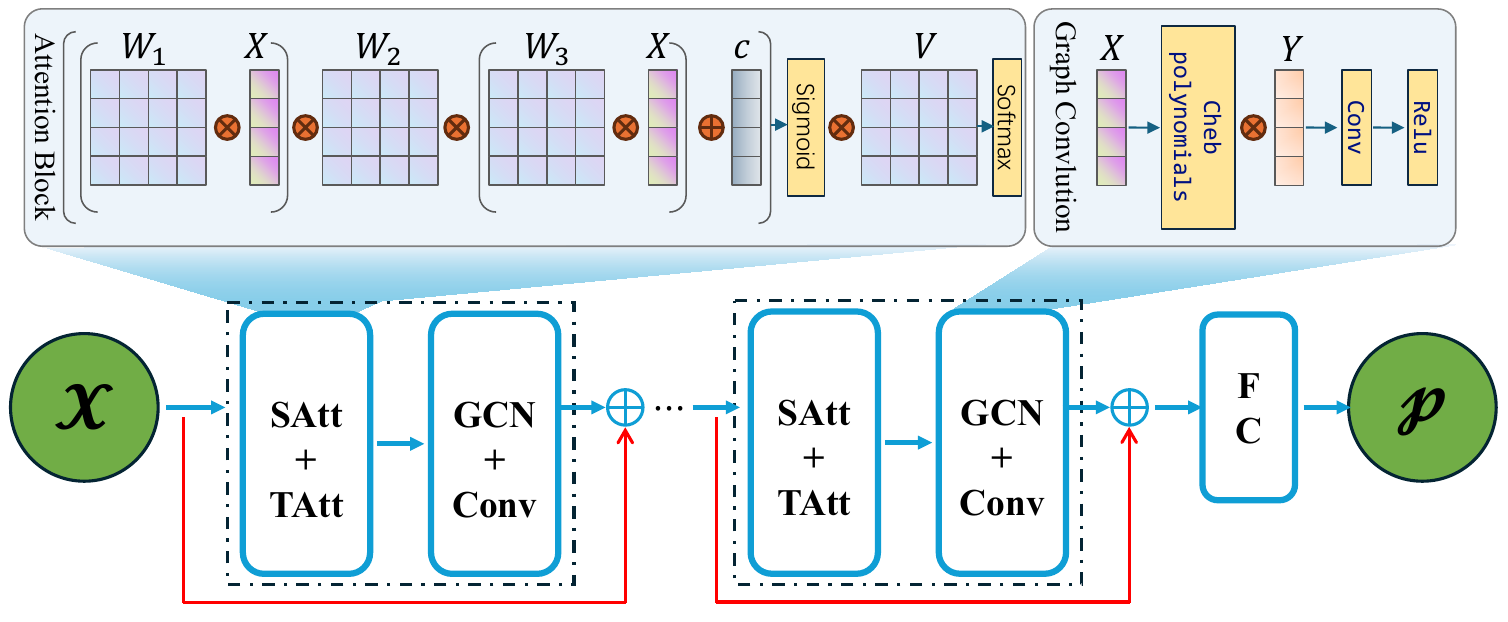}
    \caption{Popularity prediction network structure. $\mathcal{X}$ and $\mathcal{P}$ denote the historical and future popularity of various views, respectively, with red arrows indicating residual connections. }
    \label{fig:gnn}
\end{figure}

Fig. \ref{fig:gnn} shows the architecture of the popularity prediction module, which comprises two main components: spatial-temporal attention and graph convolution. Specifically, the attention part can be exemplified by the spatial attention, which includes a set of learnable parameters denoted as $\mathbf{W}_y$, $\mathbf{c}_y$, and $\mathbf{V}_y$ in the network layer.
\begin{align}
& \mathbf{Y} =   \sigma ((\mathcal{X}_\tau \mathbf{W}_{y,1}) \mathbf{W}_{y,2} (\mathbf{W}_{y,3} \mathcal{X}_{\tau})^T + \mathbf{c}_y)\mathbf{V}_y \\
& e^\mathbf{Y}_{i,k} = exp(\mathbf{Y}_{i,k}) \\
& \mathbf{Y}^{\prime}_{i,k} = \frac{e^\mathbf{Y}_{i,k}}{\sum_{k=1}^{N} e^\mathbf{Y}_{i,k}}
\end{align}
where $\mathbf{Y}^{\prime} \in \mathcal{R}^{N \times N}$ is the spatial attention matrix of graph nodes. It is obtained by applying the attention mechanism to the graph's adjacency matrix $\mathbf{A}$ and can dynamically adjust the weights between vertices during graph convolutions. 
Temporal attention is similar to spatial attention.
\begin{align}
& \mathbf{Z} = \sigma ((\mathcal{X}_\tau \mathbf{W}_{z,1}) \mathbf{W}_{z,2} (\mathbf{W}_{z,3} \mathcal{X}_{\tau})^T + \mathbf{c}_z)\mathbf{V}_z  \\
& \mathbf{Z}^{\prime}_{i,k} = \frac{e^\mathbf{Z}_{i,k}}{\sum_{k=1}^{N} e^\mathbf{Z}_{i,k}}
\end{align}
where $\mathbf{W}_z, \mathbf{c}_z, \mathbf{V}_z$ are learnable parameters. $\mathbf{Z}^{\prime}$ is the temporal attention matrix in a time slice.

Similar to \cite{guo2019attention}, we define a simplified spatial graph convolution as:
\begin{align}
g_{s} *G (\mathcal{X}_{\tau}) = \sum_{m=1}^{M} s_m (T_m(\tilde{L}) \odot \mathbf{Y}^{\prime}) \mathcal{X}_{\tau}\label{eq:sgnn}
\end{align}
where $*G$ represents a graph convolution operation(GCN). $M$ denotes the number of neighbours used in the convolution, $s$ is the convolution kernel parameter, and $T_m(\tilde{L})$ is the Chebyshev polynomial term defined on $\mathcal{X}$. 
To incorporate temporal information, we apply the temporal attention matrix $\mathbf{Z}^{\prime}$ to  $g_{s} *G (\mathcal{X}_{\tau})$, and perform a standard convolution operation in a time chunk using  $g_{t} *G$.
Multiple spatial-temporal blocks are stacked to extract a larger range of dynamic spatial-temporal correlations. Finally, a fully connected layer with a ReLU activation function is appended to ensure that the output of each component has the same dimension and shape as the forecasting target.
\begin{align}
    \mathcal{P}^{d} = \mathbf{W} \odot \operatorname{Relu}( \operatorname{FC}(g_{t} *G(\mathbf{Z}^{\prime} g_{s} *G (\mathcal{X}_{\tau}))))
\end{align}
where $\mathbf{W}$ is a learnable parameter. We exploit a mean absolute error (MAE) loss function to learn the parameters of GNN $F$.
\begin{align}
\mathcal{L}(j) = \frac{1}{N}\sum_{i=1}^{N}|\hat{p}_{i,j}-\hat{x}_{i,j}|
\end{align}
\subsection{Popularity-adaptive Bit Allocation}

With the predicted popularity of video chunks, we propose a bit allocation scheme for each representation of VARFVV. We construct the Lagrange function with the Karush-Kuhn-Tucker (KKT) conditions to solve Eq. 7, as:
\begin{align}
& L(\lambda, R_{1,j},\dots,R_{N,j},\hat{R}_{1,j},\dots,\hat{R}_{N,j}) \nonumber \\
& = \sum_{i=1}^{N} p_{i,j}log(1+R_{i,j}/\eta )+\sum_{i=1}^{N} \hat{p}_{i, j}log(1+\hat{R}_{i,j}/\hat{\eta} ) \nonumber \\ 
& - \mu_1 \cdot \mu_3\sum_{i=2}^{N}\hat{p}_{i,j}(\hat{R}_{i,j}-\hat{R}_{i-1,j})^2 \nonumber \\
& - \mu_1 \cdot \sum_{i=2}^{N}\hat{p}_{i,j}(\hat{R}_{i,j}/\hat{\eta}-R_{i-1,j}/\eta)^2 \nonumber \\
& - \mu_2\sum_{i=1}^{N}p_{i,j}(R_{i,j}-R_{i,j-1})^2 \nonumber \\
& +\lambda \left(R_j-\sum_{i=1}^{N}(\hat{R}_{i,j}+R_{i,j})\right) 
\end{align}
where $\lambda$ is the Lagrange multiplier, and $R_j$ is the target number of bits for $j$-th time slot. 
\begin{equation}
\begin{aligned}
 & R_j=\frac{R_{avg}\times(N_{coded}+SW)-R_{coded}}{SW} \\
 & R_{avg}=R_{tar}\times T_d
\end{aligned} 
\label{target}
\end{equation}
where $R_{tar}$ is the target bitrate for all representations of VARFVV, and $T_d$ is the time duration of a video chunk. $N_{coded}$ and $R_{coded}$ denote the number of the coded time slots and bit cost for all the coded representations, respectively. $SW$ is the size of sliding window, which aims to make bitrate adjustment smoothly. Then we let $\frac{\partial L}{\partial {R}_{i,j}}=0$, $\frac{\partial L}{\partial \hat{R}_{i,j}}=0$. 
\begin{equation}
\begin{aligned}
\frac{\partial L}{\partial {R}_{i,j}} = & \frac{p_{i,j}/\eta }{1+R_{i,j}/\eta}  -2 \mu_2 p_{i,j} (R_{i,j}-R_{i,j-1}) \\ & -\lambda =0
\end{aligned} 
\label{r}
\end{equation}
\begin{equation}
\begin{aligned}
 & \frac{\partial L}{\partial \hat{R}_{i,j}}  = \frac{\hat{p}_{i,j}/\hat{\eta} }{1+\hat{R}_{i,j}/\hat{\eta} } - 2 \mu_1\mu_3 \hat{p}_{i,j} (\hat{R}_{i,j}-\hat{R}_{i-1,j})  \\ 
  & - 2 \mu_1  \hat{p}_{i,j} (\hat{R}_{i,j}/\hat{\eta}-R_{i-1,j}/\eta) / \hat{\eta}-\lambda =0 \\
\end{aligned}
\label{hr}
\end{equation}

We use the dichotomy methodology to obtain the accurate value of 
 $\lambda$, $R_{i,j}$ and $\hat{R}_{i,j}$, as shown in Algorithm \ref{alg:sample}. Our bit allocation strategy dynamically allocates bits to each video chunk based on its popularity. The perceptually less popular views are compromised to save more bits for the views with more perceptual relevance. We can thus achieve a higher visual quality of FVV under the same target bits condition. 

\begin{algorithm}[t]
\caption{Popularity-adaptive Bit Allocation}
\label{alg:sample}
\KwIn{$\lambda_{min}$, $\lambda_{max}$, $R_j$, $\eta$, $\hat{\eta}$, $\epsilon$}
\KwOut{$\lambda$, $R_{i,j}$, $\hat{R}_{i,j}$}
\ 1:  \quad \textbf{while} \emph{Ture} \textbf{do}

\ 2: \quad  \quad  $\lambda = (\lambda_{max} + \lambda_{min}) / 2$ \
    
\ 3: \quad   \quad \textbf{for} \emph{$i$ = 1 : N} \textbf{do}
    
    {
\ 4: \quad  \quad \quad  Derive the view popularity $p_{i,j}$, $\hat{p}_{i,j}$\
      }

\ 5: \quad  \quad \textbf{end for}

     
\ 6: \quad  \quad  Calculate $R_{j}$ by Eq. \ref{target} \
    
\ 7: \quad   \quad \textbf{for} \emph{$i$ = 1 : N} \textbf{do}
    
    {
\ 8:  \quad     \quad \quad Calculate $R_{i,j}$, $\hat{R}_{i,j}$ by solving Eq. \ref{r}, \ref{hr} \
    }
    
\ 9: \quad  \quad \textbf{end for}

10: \quad  \quad \textbf{if} $\sum_{i=1}^{N}(\hat{R}_{i,j}+R_{i,j}) > R_{j}$ \textbf{then} 
   
 11:  \quad \quad \quad $\lambda_{min} = \lambda$\ 
   
 12:  \quad \quad \textbf{else} 
    
  13:  \quad \quad \quad $\lambda_{max} = \lambda$ 
     
  14:  \quad \quad \textbf{end if}\

  15:  \quad \quad Iteration until
  
  16: \quad \quad \quad $-\epsilon R_{j} < R_{j} - \sum_{i=1}^{N}(\hat{R}_{i,j}+R_{i,j}) < \epsilon R_{j}$
  
  17: \quad \quad \quad or achieve max iteration number

  18: \quad \textbf{end while}
    
\end{algorithm}

\section{System Implementation}
\label{sec:imple}

\subsection{Capturing and Adaptive Encoding}
\begin{figure}[tbp]
  \centering
  \includegraphics[width=\linewidth]{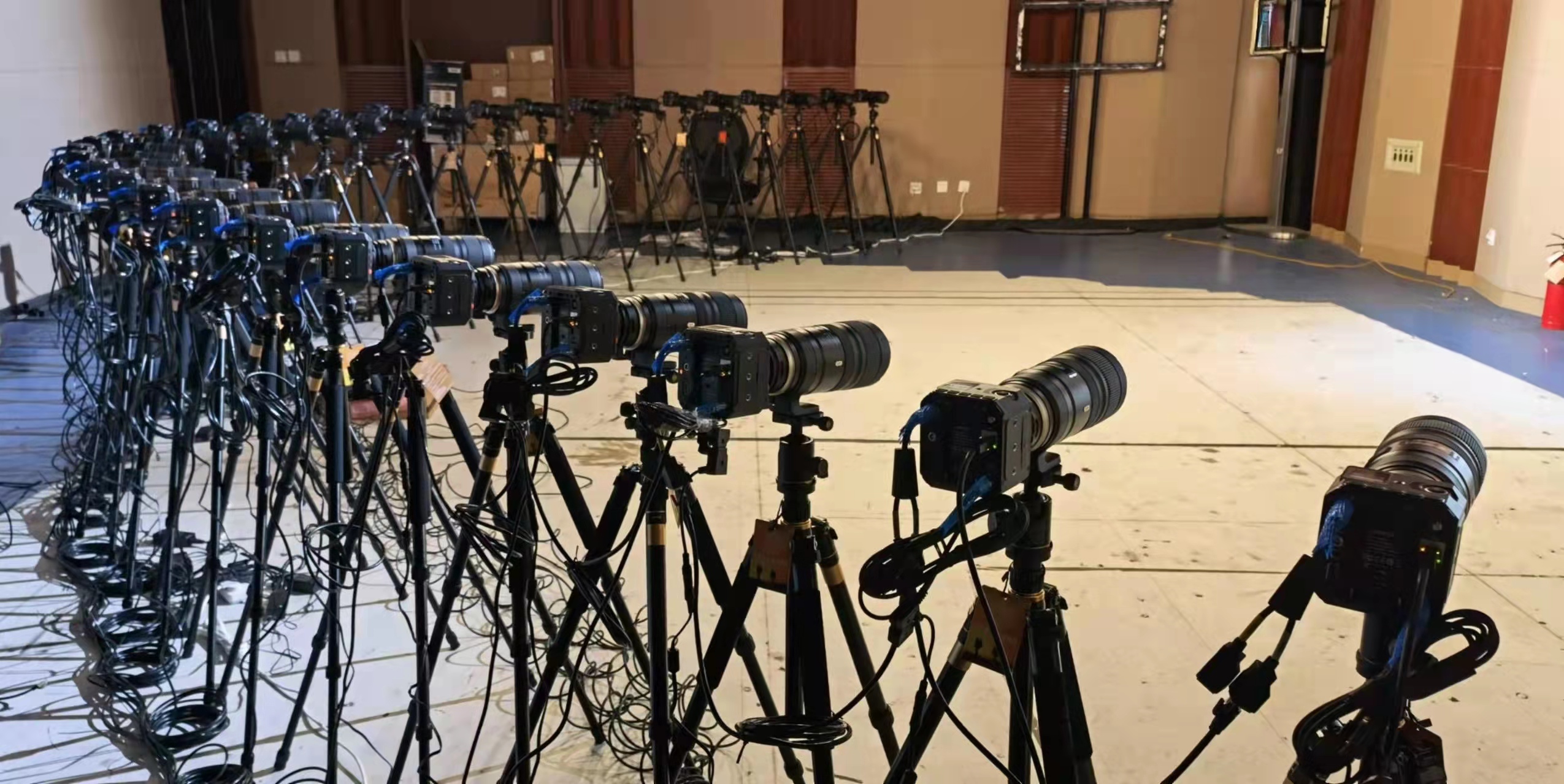}
  \caption{A set of multiview capturing system.}
  \label{capturing}
\end{figure}
As shown in Fig.\ref{capturing}, we adopt a large array of time-synchronized and closely spaced ZCAM cameras to capture the same dynamic 3D scene from $N$ different views. Each camera is equipped with a hardware synchronization unit that can trigger all cameras to shoot simultaneously. After the cameras are installed, they are calibrated to determine the exact camera parameters, including internal and external parameters. To provide a seamless visual experience for clients switching between views during the FVV live-streaming service, we utilize CUDA through a C++ interface to calibrate the images at the outset.

We implement ``\textbf{Multiview Encoder}" using the FFmpeg library, specifically utilizing the H.264 encoder. The corrected pictures for each view are encoded into \emph{view-switching representation} and \emph{view-constant representation}. Each representation of the stream records a PTS for each frame, which is calculated based on the acquisition time of that frame. The choice of the H.264 encoder balances compression efficiency and computational complexity, ensuring the overall real-time performance and stability of the FVV system.

Meanwhile, we collect users' historical trajectories. We record the number of times each view is viewed by users at each video chunk. The popularity of each view is calculated according to the number of times each view is viewed by users in each video chunk. The cloud server predicts view popularity on view-constant and view-switch representations. According to the optimal solution, VARFVV optimizes the bit allocation of each view and then updates the coding configurations. These multiview video streams are encoded and transmitted to the edge server in real-time. At this time, the bits contained in video chunks from different views are different due to the different popularity of users' viewing, and more bits are allocated to video chunks viewed by more users.
 
\subsection{Synchronization and Frame Reassembly}

The transmission speed of multiview video streams in the network is fluctuating,  which results in variability in the time that each video stream arrives at the edge server. To avoid inconsistencies between views when switching views, we implement a real-time synchronization method that achieves high tolerance for time differences between different video streams.

We use the FFmpeg library to demultiplex the received live streams and obtain the PTS of each frame. We then synchronize frames in multiview video streams according to PTS. The specific process is as follows: (i) Create sync buffers for each live stream to receive video frames. (ii) Start a thread to synchronize the frames in the buffers based on the PTS. The frames in the sync buffers are shared by all users. We select the desired frames from the sync buffers based on each user's view trajectory and current PTS, and reassemble them into a live stream. For example, when a user's view remains unchanged, the edge selects time-consecutive frames from the \emph{view-constant representation} of that view and reassembles them into a live stream for the user. When a user switches views, the edge selects frames from the \emph{view-switching representation} corresponding to the user's viewing track and reassembles them into the live stream. As we only demultiplex and reassemble video streams instead of transcoding, the increase in computational complexity is almost negligible.

\subsection{Interaction and Distribution}

We implement a signaling server to enable users to establish a WebRTC connection between edge and client. When a client sends a request to the signaling server to experience the FVV service, the signaling server assigns an edge server with service resources to the client. Then a connection is established through a discovery and negotiation process. The client can request video data from the edge server through interactive operations such as play and swipe. For example, when the client swipes to the left, it sends an interaction signal to the corresponding edge server, which generates an FVV stream consistent with the user's selected trajectory in both time and view.  The generated FVV stream is delivered to the client via  WebRTC. Our approach extracts and transmits the desired frames based on user-selected trajectories, which effectively reduces the transmission bandwidth.

\section{Experimental Results}

\subsection{Training}

\begin{figure}[t]
    \centering
    \includegraphics[width=\linewidth]{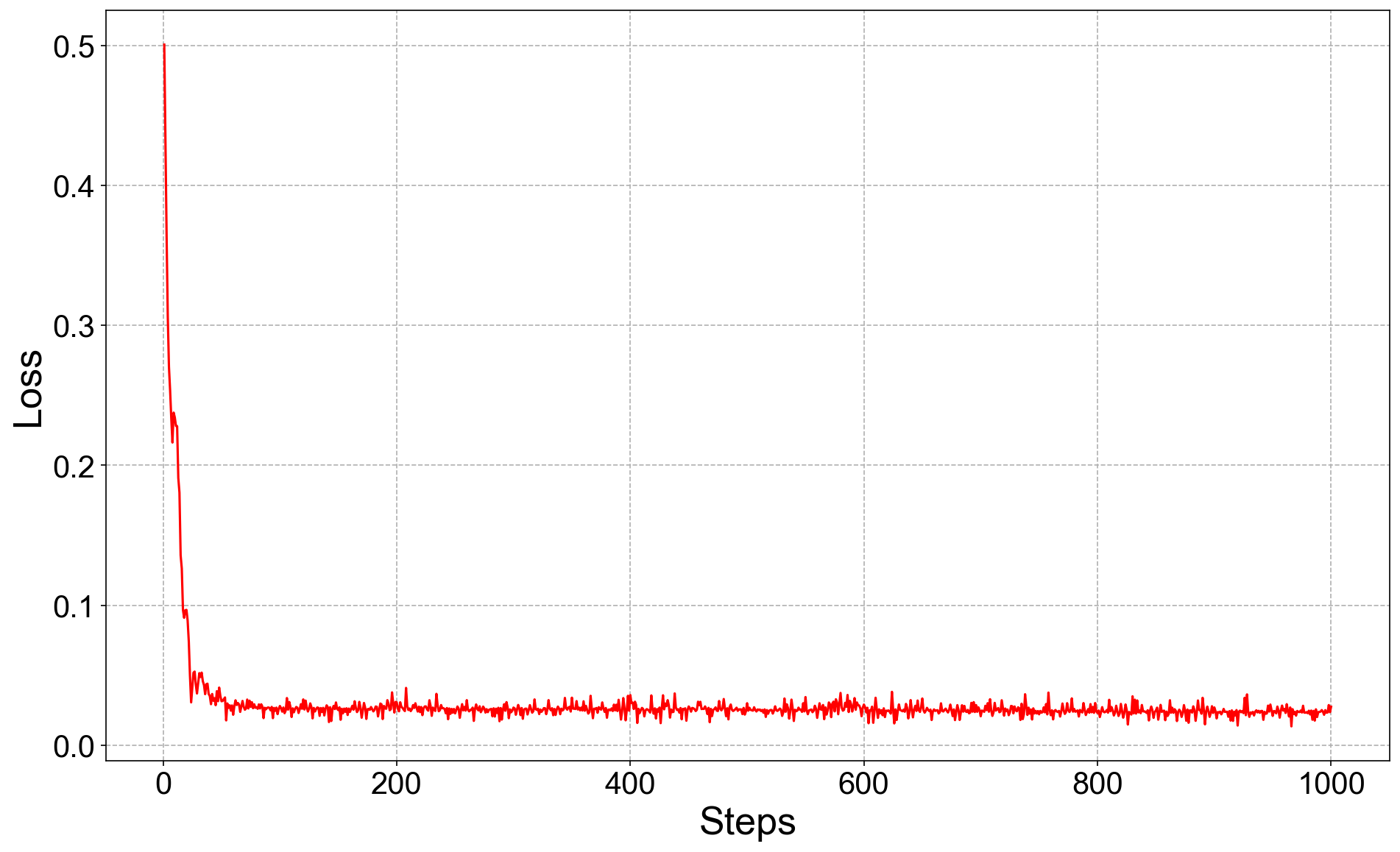}
    \caption{Loss curve for training process.}
  \label{train_loss}
\end{figure}

To evaluate the performance of VARFVV, we construct an FVV dataset comprising 330 videos from 10 scenes, including basketball, opera, martial arts, etc. We recruit 82 participants to use our VARFVV to freely view these videos and record their view traces. A cloud server with an AMD Ryzen 7 3700X CPU@3.60 GHz and an NVIDIA 2080ti GPU is used for popularity training and prediction. Our view popularity prediction method combines PPC and GNN, where PPC uses ${x}_{i,j-1}$ as ${p}_{i,j}$, while GNN is trained online with historical data samples of $\hat{x}_{i,j}$ during the initial 10 seconds. We configure the GNN training with $M=2$, utilizing an MAE loss function alongside a learning rate of $0.005$ and a batch size of 32. GNN undergoes 50 epochs during the initial training to achieve optimal results.

After the initial training, the obtained weights are used to predict the popularity distribution of users. During the online learning phase, each new data frame is immediately added to the training set, and the GNN model parameters are updated in real-time, ensuring continuous adaptation to the latest data.
The analysis of the training process, as illustrated in Fig.\ref{train_loss}, demonstrates a consistent reduction in the MAE loss of the GNN as the number of steps increases. Notably, the loss function converges and approaches zero after only a few steps.

Tab. \ref{tab:time} provides insights into the training and prediction times. The training time reflects the duration to train 50 epochs on 10 seconds of data, averaging 1.61s. This indicates a swift training and update process. The average prediction time is 0.16ms, showcasing the model's real-time capability in predicting popularity distribution results.

\begin{figure}[t]
    \centering
    \includegraphics[width=0.9\linewidth]{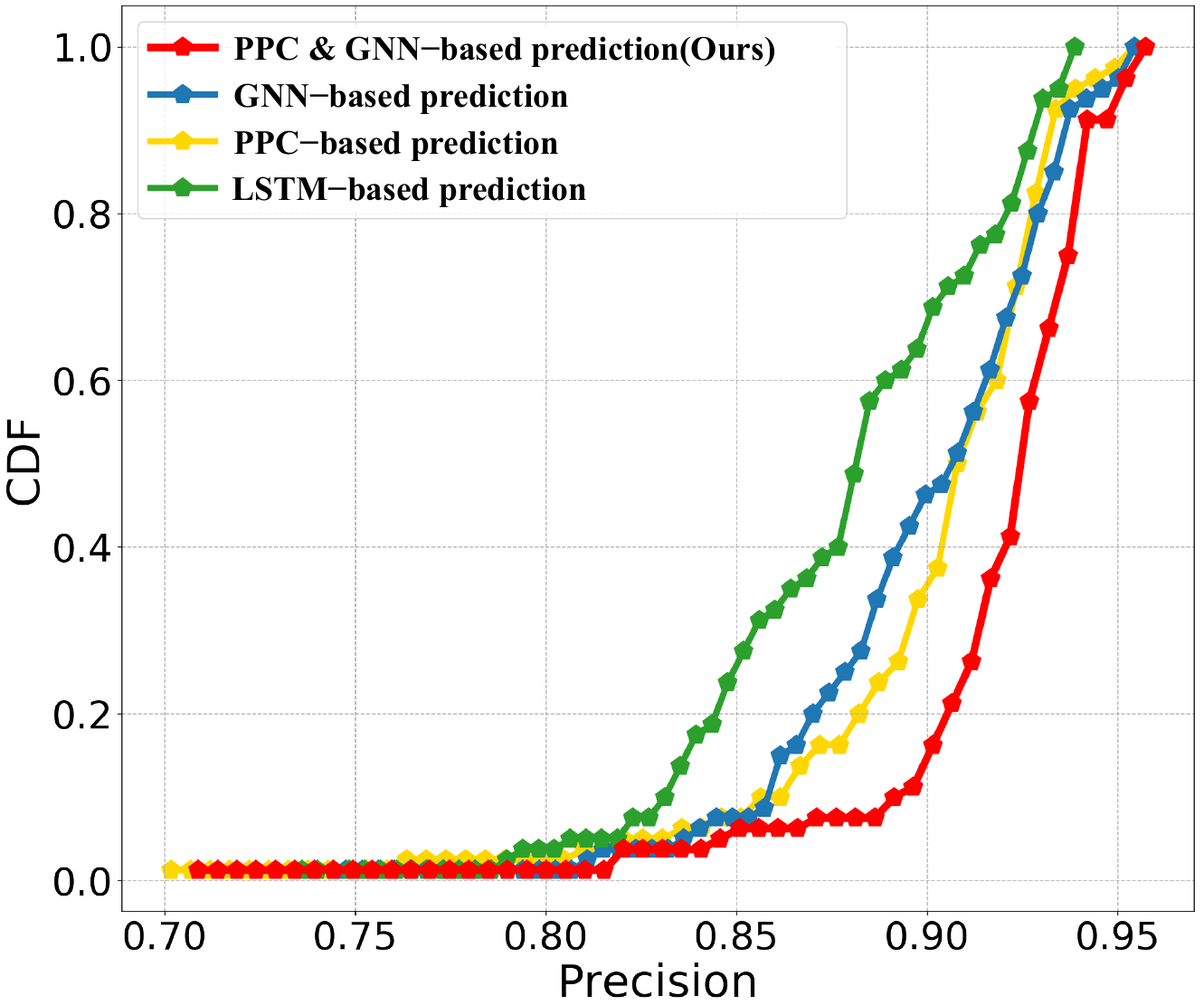}
    \caption{The CDF of popularity prediction precision on the dataset where the length of a video chunk is one second.}
  \label{precision}
\end{figure}

\begin{table}
  \centering
  \caption{Training time and prediction time}
  \label{tab:time}
  \begin{tabular}{cccc}
    \toprule
    Scene & View & Training Time & Prediction Time \\
    \midrule
    Basketball & 23 & 1.49s & 0.15ms \\
    Opera & 48 & 1.56s & 0.19ms \\
    Martial Arts & 48 & 1.87s & 0.15ms \\
    Teaching & 13 & 1.52s & 0.13ms \\
    Average & - & 1.61s & 0.16ms \\
    \bottomrule
  \end{tabular}
\end{table}

\subsection{Evaluation}

 \textbf{Precision of View Popularity Prediction.} We evaluate our view popularity prediction method in comparison to an LSTM-based prediction method \cite{9343267}. We also conduct an ablation study to individually evaluate the performance of the proposed PPC and GNN-based view popularity predictions. Inspired by \cite{8737361}, we use the precision of popularity prediction to investigate the performance.
\begin{align}
    & Precision(j) = 1- \nonumber \\
    & \sqrt{\frac{\sum_{i=1}^{N} (p_{i,j} - x_{i,j})^2+\sum_{i=1}^{N} (\hat{p}_{i,j} -\hat{x}_{i,j})^2}{2N}}.  \label{ev_pre2} 
\end{align}

Fig. \ref{precision} demonstrates the cumulative distribution function (CDF) of precision in predicting popularity. Our popularity prediction scheme outperforms the other three schemes, primarily due to its ability to model both view-constant and view-switching scenarios. Specifically,  PCC is effective for view-constant scenarios by leveraging the popularity of previous chunks, but struggles with view-switching due to its inability to capture dynamic user behavior changes. LSTM captures temporal dynamics in view-switching scenarios but fails to account for spatial relationships between views, limiting its performance. In contrast, GNN captures both spatial correlations and temporal dynamics through a graph-based representation, enabling it to better adapt to user focus shifts during view switching and achieve higher accuracy. This also aligns with the results in Fig. \ref{fig:GNNLSTM}. However, GNN lacks explicit optimization for view-constant scenarios. Our method combines PCC for constant views and GNN for dynamic views, ensuring superior prediction accuracy across both contexts.

\begin{figure}
    \centering
    \includegraphics[width=0.9\linewidth]{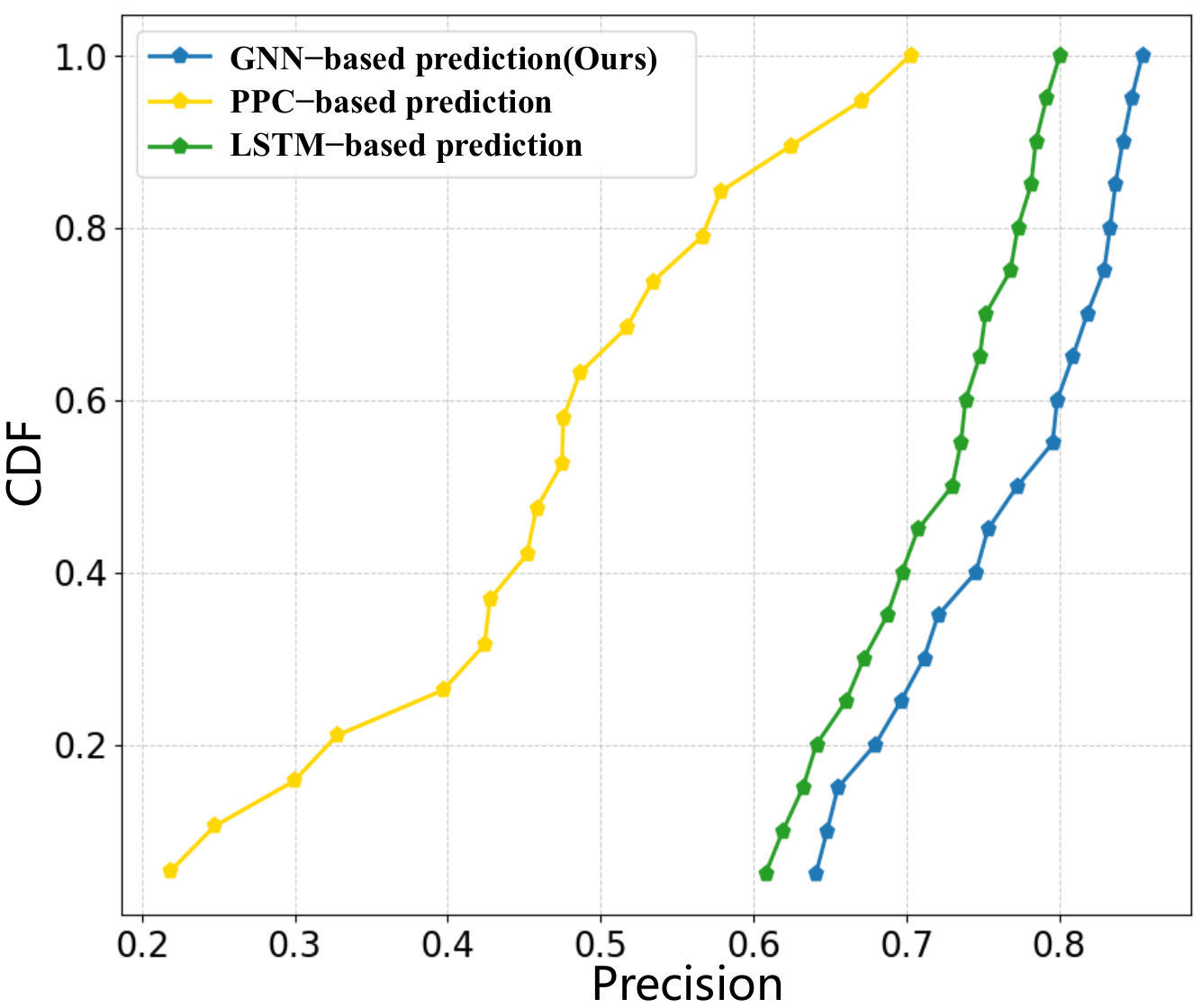}
    \caption{The CDF of popularity prediction accuracy under sudden changes in user
behavior during view switching.}
    \label{fig:suddenchanges}
\end{figure}

We also evaluate the performance of view popularity prediction under continuous and sudden changes in user behavior during view switching. For this experiment, we use a 20-second sequence from the opera scenario, where users continuously switch views, simulating dynamic changes in behavior.  As shown in Fig. \ref{fig:suddenchanges}, our GNN-based prediction maintains competitive accuracy even under these challenging conditions, outperforming both PPC and LSTM. This is due to its ability to model spatial correlations between views and capture the dynamic changes in user preferences during view switching. However, as expected, the prediction accuracy for stable interactions (shown in Fig. \ref{precision}) is higher than that for scenarios with sudden changes, due to the increased variability introduced by rapid shifts in user behavior.

\begin{figure}[t]
  \centering
  \subfigure[The total number of bits per second is $10\cdot N$ Mbps ($N=23$). ]{
    \label{fig:subfig:qoe1} 
    \includegraphics[width=0.9\linewidth]{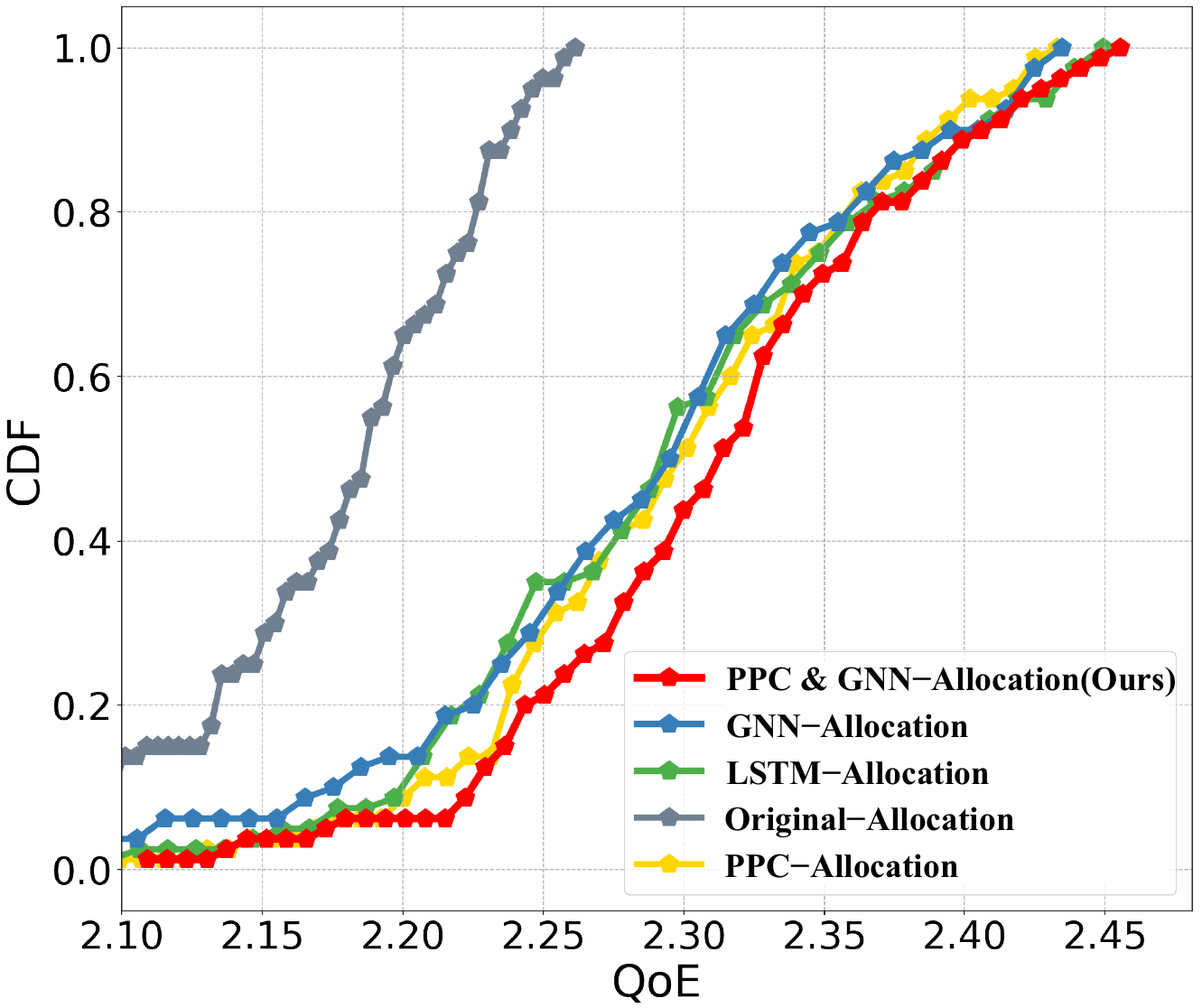}}
  \subfigure[The total number of bits per second is $20\cdot N$ Mbps ($N=23$).]{
    \label{fig:subfig:qoe2} 
    \includegraphics[width=0.9\linewidth]{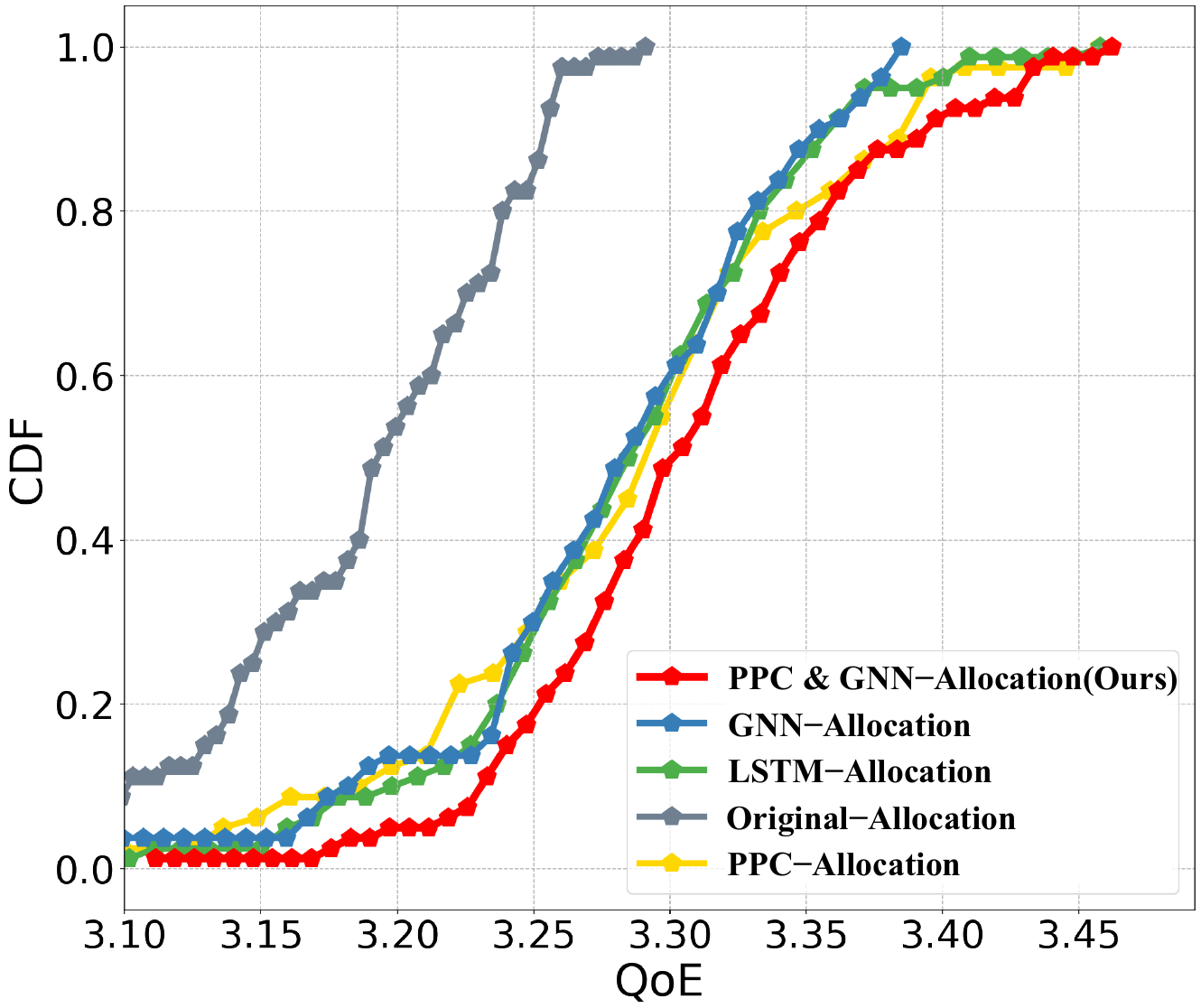}}
  \caption{The CDF of user QoE on the dataset where the length of a video chunk is one second.}
  \label{qoe_4} 
\end{figure}

\noindent{\textbf{User QoE. }} We evaluate the proposed bit allocation strategy against four alternatives: original-allocation, PPC-allocation, LSTM-allocation, and GNN-allocation. The original-allocation strategy distributes equal bits for all the views, while the other methods implement the popularity-adaptive bit allocation using different popularity prediction algorithms. For our method, we specify the tuning parameters for user QoE: $\eta = 1, \hat{\eta} = 4, \mu_1 = 1, \mu_3 = 1, \mu_2 = 1/16, \epsilon = 0.005, \lambda_{min} = 0, \lambda_{max} = 100$. By applying these parameters as described in Algorithm \ref{alg:sample}, we solve for the optimal $ R_{i,j} $ and $ \hat{R}_{i,j}$.

Fig. \ref{qoe_4} presents the measured user QoE for each scheme, where each scheme utilizes the same total number of bits across all representations. We observe that our proposed bit allocation scheme achieves the best QoE compared to the others, both in high and low bit scenarios, validating the effectiveness of the proposed approach. The original-allocation method consistently performs worse than the others because the popularity-adaptive bit allocation allows us to borrow bits more aggressively from less popular views and allocate them to more popular views. Additionally, the proposed bit allocation scheme guarantees a higher minimum QoE than the other schemes, indicating its ability to maintain a minimum level of user satisfaction when watching free-view videos. These results further demonstrate the efficacy of our proposed popularity prediction model, which is identical to the results presented in Fig. \ref{precision}.

  \begin{figure}[t]
    \centering
    \includegraphics[width=0.9\linewidth]{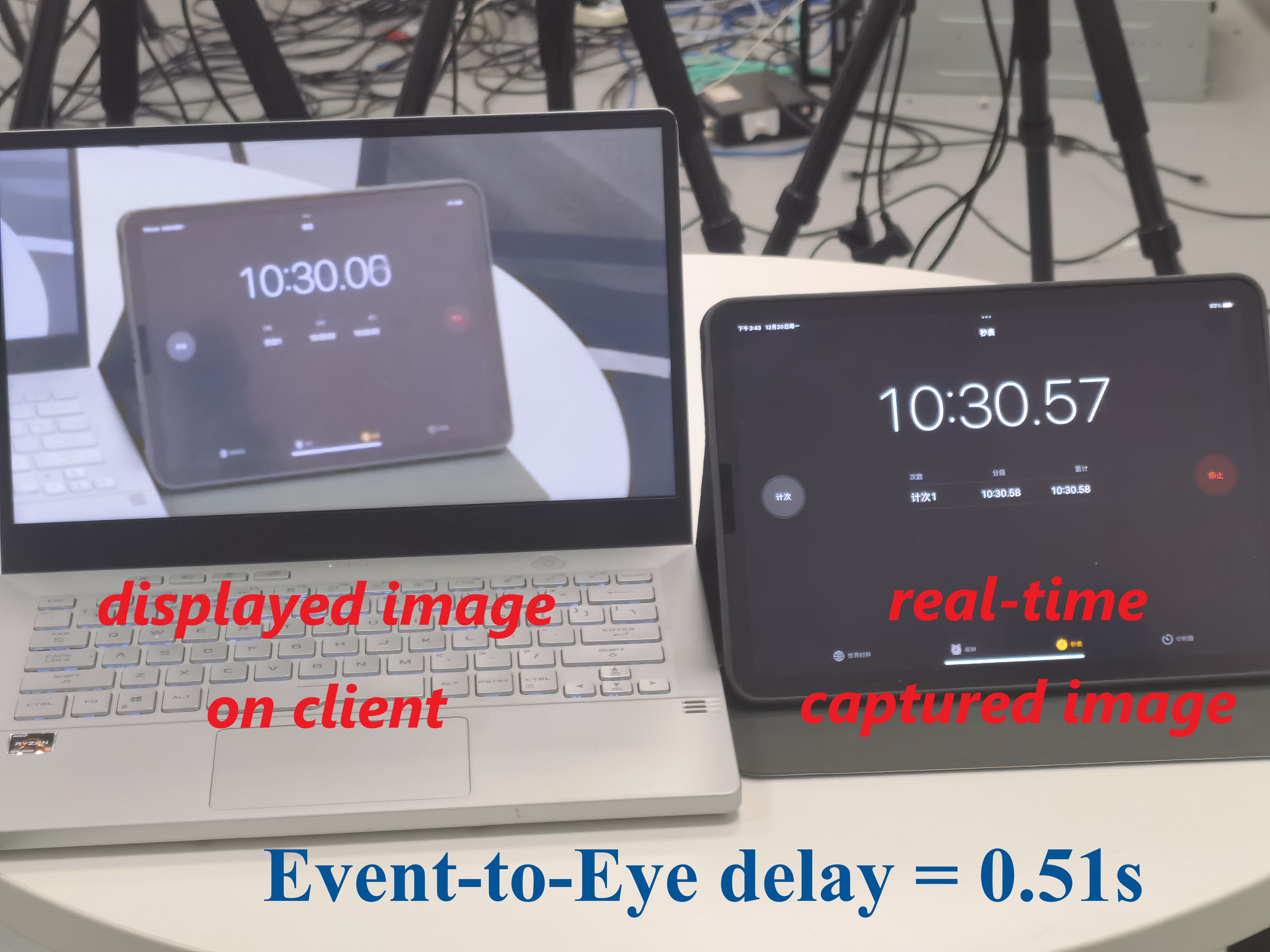}
    \caption{Event-to-eye delay visualization in VARFVV. This figure illustrates the delay comparison between the VARFVV client device and the capture side, highlighting the timestamps used to calculate the event-to-eye delay.}
  \label{delay}
\end{figure}

 \begin{figure*}[t]
\includegraphics[width=0.99\linewidth]{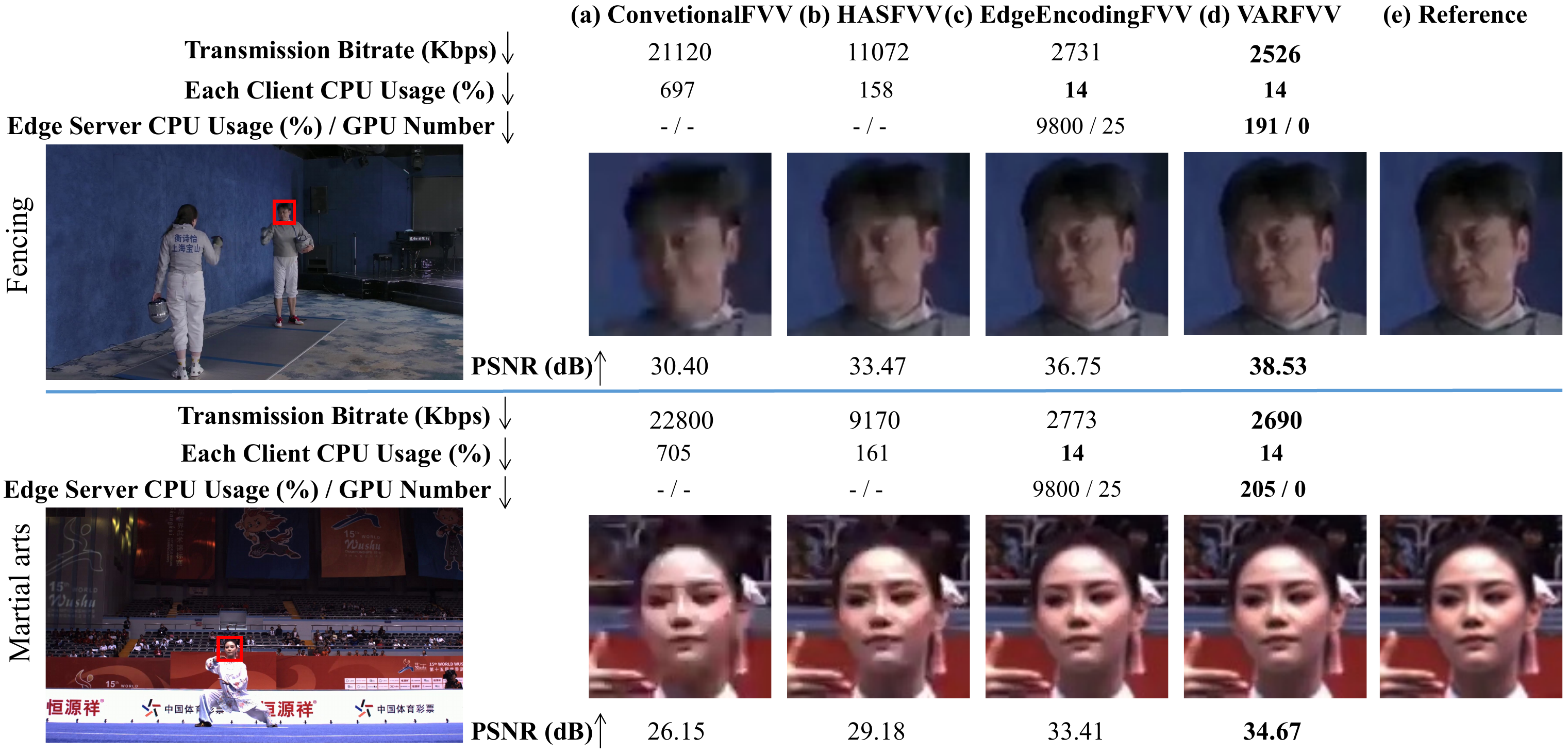}
\caption{Comparison of our VARFVV with ConventionalFVV \cite{fvtv1,fvtv2}, HASFVV \cite{8325530}, and EdgeEncodingFVV \cite{9431586} in the ``Low-interactivity Scenario".  The number of views of the ``Fencing" scene and the ``Martial Arts" scene is 48. (Simulate 500 users.)}
\label{fix4}
\end{figure*}

 \noindent{\textbf{Delay.}} We assess three critical delay metrics that impact the FVV viewing experience: start-up delay, view-switching delay, and event-to-eye delay. The start-up delay is the time taken from sending a streaming request to the appearance of the first video frame on the screen \cite{delay10.1145/3394171.3413539}. The view-switching delay refers to the duration between requesting a view switch and the display of the desired view \cite{8325530}. Event-to-eye delay measures the time for an event, captured by the camera, to be displayed on the client's screen \cite{delay10.1145/3394171.3413539}.

\begin{figure*}[h]
\includegraphics[width=1\linewidth]{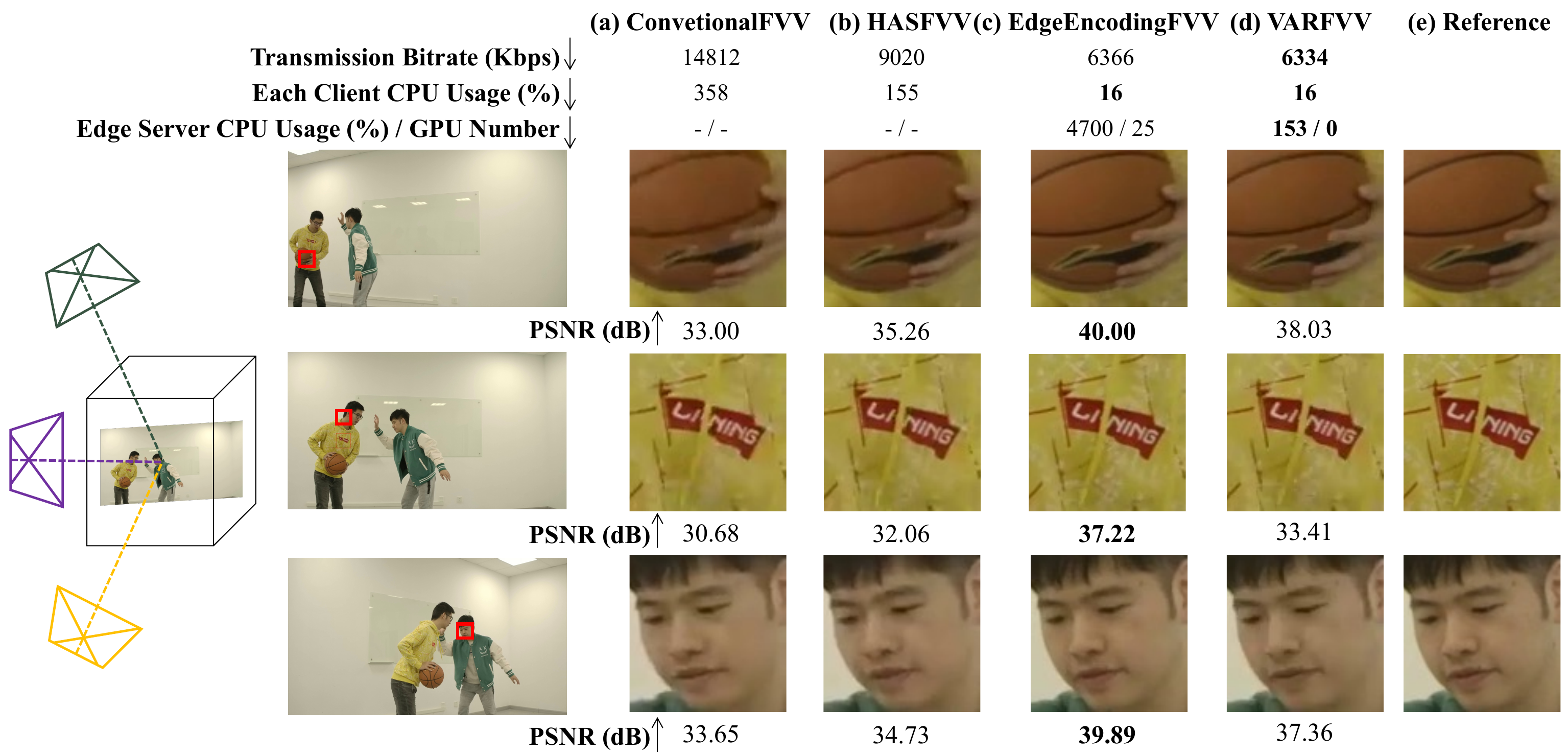}
\caption{Comparison of our VARFVV with ConventionalFVV  \cite{fvtv1,fvtv2}, HASFVV \cite{8325530}, and EdgeEncodingFVV \cite{9431586} in the ``High-interactivity Scenario". The number of views of ``Basketball" scene is 23. (Simulate 500 users.)}
\label{change4}
\end{figure*}
 
 To accurately measure these delays, we record the intervals between the requests and significant changes in the pixel values of the frames. Our analysis revealed that the average start-up delay in the VARFVV is approximately 195.1 ms, while the view-switching delay is notably shorter at about 71.5 ms. Notably, this view-switching delay is typically too brief to be noticed by the human eye. As illustrated in Fig. \ref{delay}, the event-to-eye delay is around 0.51s, as confirmed by the timestamps: the VARFVV client shows 10:30.06, while the capture side records 10:30.57. These results demonstrate VARFVV’s low-latency performance, which is critical for delivering an immersive user experience.


\subsection{Comparison}
We compare the performance of VARFVV with ConventionalFVV \cite{fvtv1,fvtv2} (receiving all views at the client), HASFVV \cite{8325530} (receiving HTTP adaptive streaming with 10 views at the client), and EdgeEncodingFVV \cite{9431586} (transcoding frames at the edge server). The performance is evaluated in terms of PSNR, transmission bitrate, and computational resources including client CPU, edge server CPU, and GPU usage. The GPU usage refers to the GPU number required. 
We simulate 500 users experiencing FVV concurrently to assess edge server computational resources. 
For a fair comparison, we maintain identical transmission bitrate for VARFVV and EdgeEncodingFVV. However, as ConventionalFVV and HASFVV require clients to receive a larger number of video streams, the bitrate per stream is reduced accordingly. In addition, EdgeEncodingFVV requires frame transcoding and therefore we configure each edge server with an AMD Ryzen 7 3700X CPU@3.60 GHz and two RTX4000 GPUs.
All methods are evaluated using the same hardware and software configurations for edge servers and clients. We categorize scenarios based on user interactivity levels: (i) ``Low-interactivity Scenario": users' navigation windows are almost constant. (ii) ``High-interactivity Scenario": users keep switching views \cite{8325530}.

Fig. \ref{fix4} presents the results of our VARFVV compared to other methods in the ``Low Interactive Scenario" at 1080p@25FPS. Unlike ConventionalFVV and HASFVV, which require high bandwidth and substantial CPU usage to receive and decode multiple streams, our VARFVV achieves superior visual quality with lower transmission bitrate and reduced client CPU usage. Furthermore, compared to EdgeEncodingFVV, our approach demonstrates higher visual quality. Since we reassemble video frames instead of transcoding them at the edge, maintaining the original video quality without imposing heavy computational demands.

Fig. \ref{change4} illustrates the results of our VARFVV compared to other methods in the ``High Interactive Scenario" at 1080p@25FPS. Our VARFVV achieves much better visual results than ConventionalFVV and HASFVV with lower transmission bitrate and lower client CPU usage. Despite exhibiting a lower PSNR than EdgeEncodingFVV, VARFVV maintains comparable subjective quality, especially during rapid view switching, where human perception is less sensitive to distortions. However, EdgeEncodingFVV demands substantially higher computational resources than VARFVV to achieve similar visual outcomes. Specifically, our single edge server does not require GPUs, whereas EdgeEncodingFVV's edge servers need 25 RTX4000 GPU cards, each capable of encoding up to 20 videos at 1080p@25fps, to serve 500 users. Furthermore, VARFVV proves more CPU-efficient, with only 153\% edge server CPU usage compared to EdgeEncodingFVV's 4700\%, primarily utilized for decoding tasks. Leveraging the continuous reassembly of multiview video frames without edge-side decoding and encoding enables each server to efficiently serve up to 500 users simultaneously, significantly reducing computational demands.

\begin{table}[t]
\caption{Quantitative Comparison When 500 Users Use an FVV Service at 4K Resolution with 23 Wiews. }
\footnotesize
\newcommand{\tabincell}[2]{\begin{tabular}{@{}#1@{}}#2\end{tabular}} 
  \centering
\resizebox{\linewidth}{!}{ 
\begin{tabular}{cccl}
\toprule
{Method} & \tabincell{c}{Each Client CPU \\ Usage (\%) $\downarrow$}   & \tabincell{c}{Edge Server CPU Usage (\%) \\ / GPU Number $\downarrow$} \\ \midrule               
{HASFVV}          & {494}         & {- / -}                 \\
{EdgeEncodingFVV} & {45} & {52560 / 100}               \\
{VARFVV}          & {\textbf{45}} & {\textbf{232} / \textbf{0}}                 \\ \bottomrule
\end{tabular}}%
  \label{tab:4kusage}
\end{table}

Furthermore, we conduct experiments to evaluate computational costs on videos at 4K@25FPS when 500 users use our VARFVV, HASFVV, and EdgeEncodingFVV. Tab. \ref{tab:4kusage} shows the results. On a client CPU, our VARFVV, as well as EdgeEncodingFVV, consumes 45$\%$ while HASFVV uses 494$\%$ because HASFVV does not exploit an edge server. On edge server, our technique occupies 232$\%$ without GPU cards whereas EdgeEncodingFVV uses 52560$\%$ with 100 GPU cards. These results show that our VARFVV is lightweight and computation-efficient on both clients and edge servers.

\section{Conclusion}

We propose and implement VARFVV, a novel FVV streaming system that achieves low switching delay and high QoE while maintaining low-cost computation and transmission. Our computationally efficient FVV stream generation approach significantly reduces the computational load at the edge by demultiplexing and reassembling video frames, making it ideal for large-scale, mobile-based UHD FVV experiences. Furthermore, we also design a PPC and GNN-based popularity prediction algorithm and a popularity-adaptive bit allocation algorithm for multiview coding to maximize the total QoE with a limited number of bits. Extensive experimental results indicate the effectiveness of our approach for FVV quality, computational resource, and transmission bandwidth, which compares favorably to the state-of-the-art.

\bibliographystyle{IEEEtran}
\bibliography{main.bbl}

\begin{IEEEbiography}[
{\includegraphics[width=1in,height=1.25in,clip]{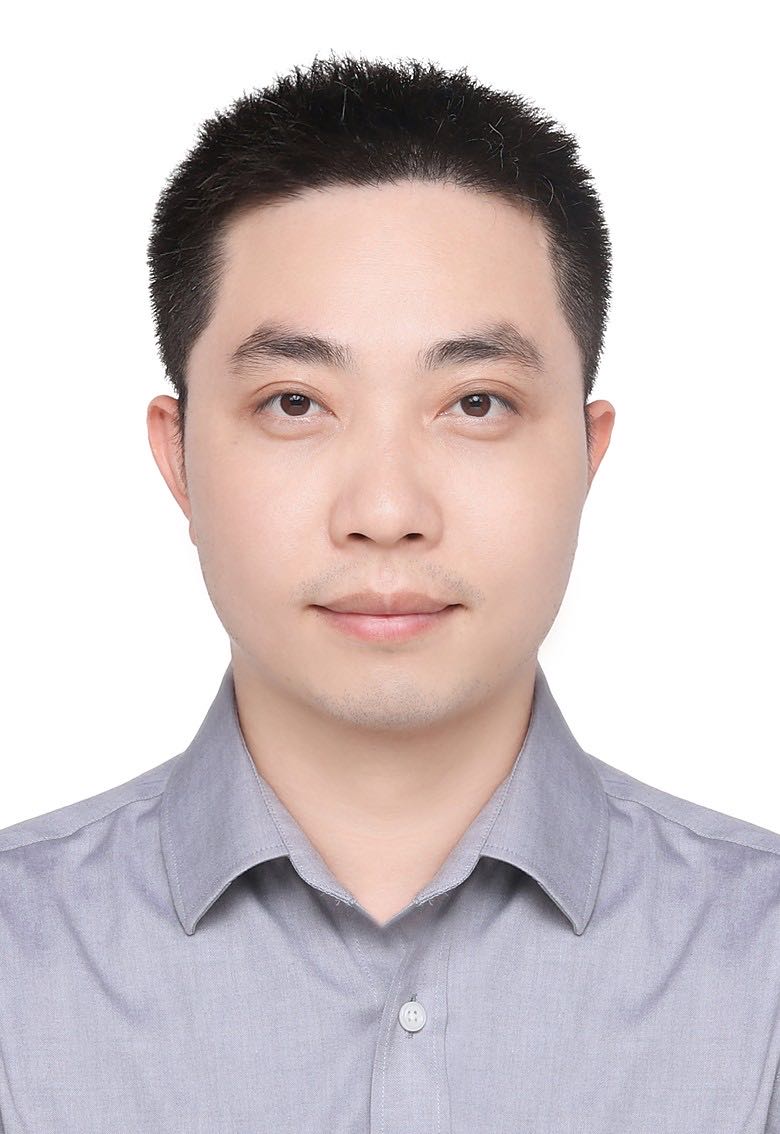}}
]{Qiang Hu} (Member, IEEE) received the B.S. degree in electronic engineering from University of Electronic Science and Technology of China in 2013, and the Ph.D. degree in information and communication engineering from Shanghai Jiao Tong University in 2019.  He is currently an Assistant Researcher at Cooperative Medianet Innovation Center, Shanghai Jiao Tong University. Before that, he was an Assistant Researcher at ShanghaiTech University from 2021 to 2023. He was a Postdoc Researcher at ShanghaiTech University from 2019 to 2021. His research interests focus on 2D/3D video compression, 3D reconstruction, and generative intelligence media. His works have been published in top-tier journals and conferences, such as IEEE Transactions on Image Processing (TIP), Conference on Computer Vision and Pattern Recognition (CVPR).
\end{IEEEbiography}

\begin{IEEEbiography}[
{\includegraphics[width=1in,height=1.25in,clip,keepaspectratio]{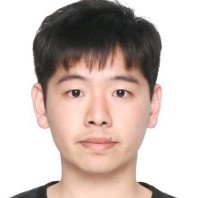}}
]{Qihan He} received his B.S. degree from Shanghai Jiao Tong University in 2020 and M.S. degree from ShanghaiTech University in 2023. His research interests lie in video transmission and compression.
\end{IEEEbiography}

\begin{IEEEbiography}[
{\includegraphics[width=1in,height=1.25in,clip,keepaspectratio]{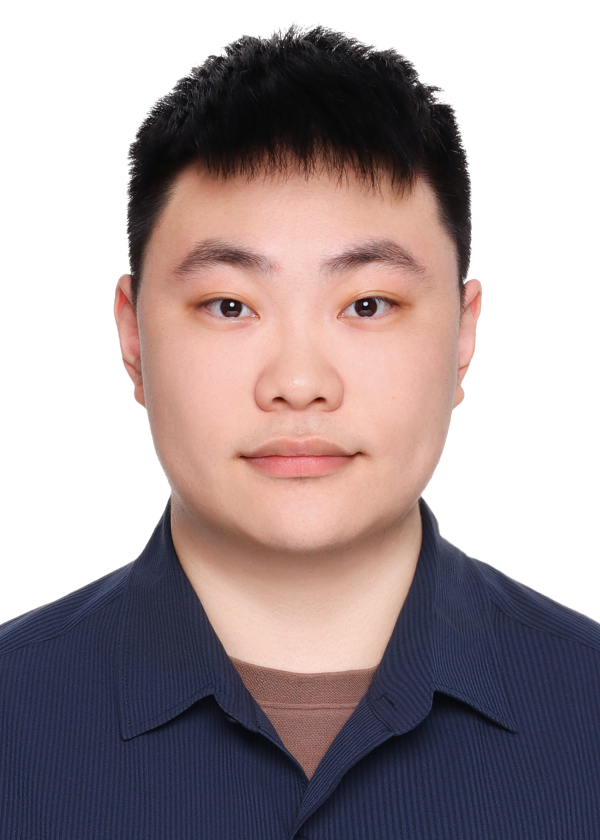}}
]{Houqiang Zhong} (Student Member, IEEE) received his B.E. degree from Shanghai University in 2020 and M.S. degree from ShanghaiTech University in 2023. He is currently pursuing a Ph.D. in the School of Electronic Information and Electrical Engineering at Shanghai Jiao Tong University. His research focuses on volumetric video and 3D reconstruction.
\end{IEEEbiography}

\begin{IEEEbiography}[
{\includegraphics[width=1in,height=1.25in,clip,keepaspectratio]{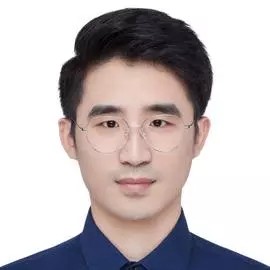}}
]{Guo Lu} ( Member, IEEE) received the B.S. degree from the Ocean University of China in 2014 and the Ph.D. degree from Shanghai Jiao Tong University in 2020. He is currently an Assistant Professor with the School of Computer Science, Beijing Institute of Technology, China. His works have been published in top-tier journals and conferences, such as IEEE Transactions on Pattern Analysis and Machine Intelligence (TPAMI), IEEE Transactions on Image Processing (TIP), Conference on Computer Vision and Pattern Recognition (CVPR), and European Conference on Computer Vision (ECCV). His research interests include image and video processing, video compression, and computer vision. He serves as a reviewer/program committee member for IEEE Transactions on Pattern Analysis and Machine Intelligence (TPAMI) and IEEE Transactions on Image Processing (TIP). He is also serving as a Guest Editor for an International Journal of Computer Vision (IJCV) Special Issue on Deep Learning for Video Analysis and Compression.
\end{IEEEbiography}

\begin{IEEEbiography}[
{\includegraphics[width=1in,height=1.25in,clip,keepaspectratio]{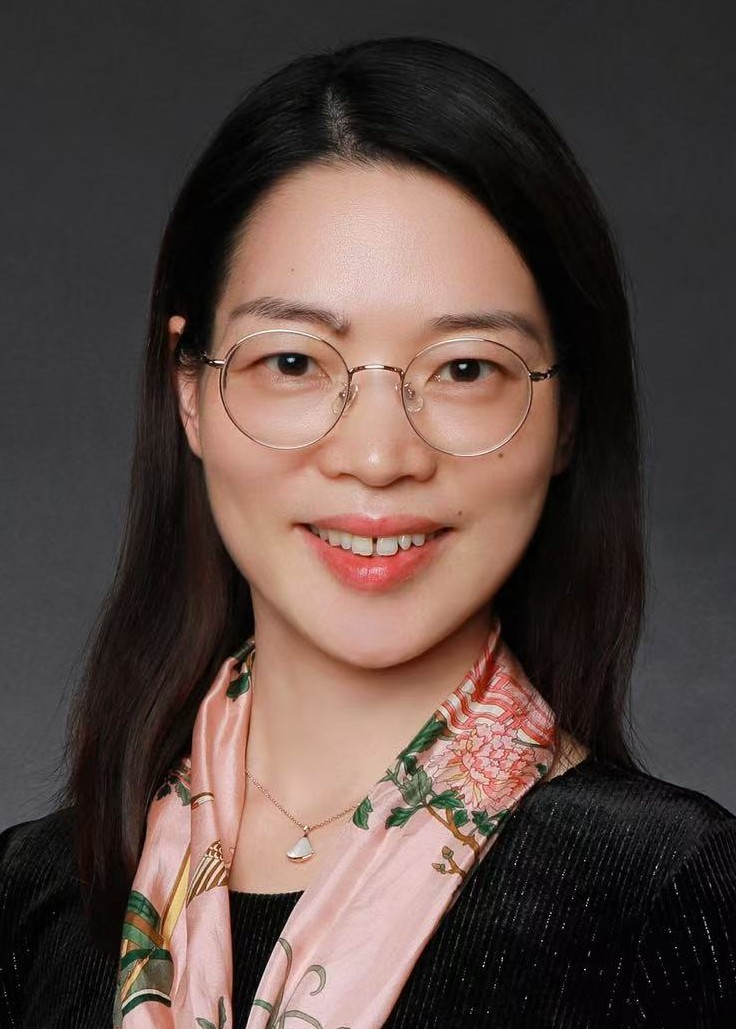}}
]{xiaoyun Zhang}( Member, IEEE), Professor of Cooperative Medianet Innovation Center (CMIC), Shanghai Jiao Tong University. She received her Ph.D (supervised by Yuncai Liu) on Pattern Recognition from Shanghai Jiao Tong University and Master (supervised by Zongben Xu) on Applied Mathematics from Xi`an Jiao Tong University respectively. She has co-authorized papers on IEEE TPAMI, TIP, CVPR, ICCV, etc., and her Ph.D. thesis has been nominated as “National 100 Best Ph.D. Theses of China”. She has been a visiting scholar of Harvard University for one year. She is also a member of State Key Laboratory of UHD video and audio production and presentation, and has collaboration with CCTV in video restoration and enhancement.
\end{IEEEbiography}

\begin{IEEEbiography}[
{\includegraphics[width=1in,height=1.25in,clip,keepaspectratio]{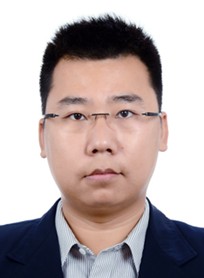}}
]{Guangtao Zhai}(Fellow, IEEE) is a professor at Department of Electronics Engineering, Shanghai Jiao Tong University. His research interests are in the fields of multimedia and perceptual signal processing. He has received the Humboldt fellowship in 2011, national PhD thesis awards of China in 2012, best student paper award of Picture Coding Symposium (PCS) 2015, best student paper award of IEEE International Conference of Multimedia and Expo (ICME) 2016, best paper award of IEEE Trans. Multimedia 2018, Saliency360! Grand Challenge of ICME 2018, best paper award of IEEE Mobile Multimedia Computing Workshop (MMC) 2019, best paper award of IEEE CVPR Dyna-Vis Workshop 2020, best paper runner-up of IEEE Trans. Multimedia 2021, 1st place of UGC VQA Contest FR Track in IEEE ICME 2021. He also received the “Eastern Scholar” and “Dawn” program professorship of Shanghai, China, NSFC excellent young researcher award and national top young researcher award in China. He is a member of IEEE CAS MSA TC and SPS IVMSP TC. He serves as Editor-in-Chief of Displays (Elsevier), he is also on the editorial board of Digital Signal Processing (Elsevier) and Science China: information Science (Springer).
\end{IEEEbiography}

\begin{IEEEbiography}[
{\includegraphics[width=1in,height=1.25in,clip,keepaspectratio]{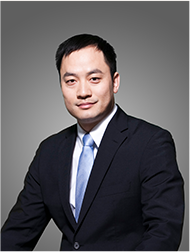}}
]{Yanfeng Wang} received the B.S. degree from PLA Information Engineering University, Beijing, China, and the M.S. and Ph.D. degrees in business management from Shanghai Jiao Tong University, Shanghai, China. He is currently the Vice Director of Cooperative Medianet Innovation Center and also the Vice Dean of the School of Electrical and Information Engineering, Shanghai Jiao Tong University. His research interest mainly include media Big Data, the emerging commercial applications of information technology, and technology transfer.
\end{IEEEbiography}

\vfill


\end{document}